\documentclass{article}
\usepackage[onecolumn]{emulateapj,epsfig}

\slugcomment{Revised version, submitted to ApJ}

\makeatletter

\newenvironment{inlinefigure}{%
\def\@captype{figure}%
\noindent\begin{minipage}{0.999\linewidth}\begin{center}}
{\end{center}\end{minipage}\smallskip}
\makeatother

\newcommand{\HII}{H{\sc ii} }

\newcommand{\Hminus}{{\rm H}^-}
\newcommand{\Htwoplus}{{\rm H}_2^+}
\newcommand{\HH}{H$_2$ }
\newcommand{\Htwo}{{\rm H}_2}

\newcommand{\simgt}{\lower.5ex\hbox{$\; \buildrel > \over \sim \;$}}

\begin{document}

\title{Formation of Primordial Stars in a $\Lambda$CDM Universe}

\author{Naoki Yoshida}
\affil{Department of Physics, Nagoya University, Furocho, Nagoya, Aichi 464-8602, Japan}
\author{Kazuyuki Omukai}
\affil{National Astronomical Observatory of Japan, 2-21-1 Osawa, Mitaka, Tokyo 181-8588, Japan}
\author{Lars Hernquist}
\affil{Harvard-Smithsonian Center for Astrophysics, 60 Garden Street, Cambridge, MA02138}
\and
\author{Tom Abel}
\affil{Kavli Institute for Particle and Astrophysical Cosmology, Stanford University, 
2575 Sand Hill Road, 
Menlo Park, CA 94025}

\begin{abstract}

Primordial stars are formed from a chemically pristine gas consisting
of hydrogen and helium. They are believed to have been born at some
early epoch in the history of the Universe and to have enriched the
interstellar medium with synthesized heavy elements before the
emergence of ordinary stellar populations.  We study the formation of
the first generation of stars in the standard cold dark matter
model. We follow the gravitational collapse and thermal evolution of
primordial gas clouds within early cosmic structures using very
high-resolution, cosmological hydrodynamic simulations.  
Our simulation achieves a dynamic range of $\sim 10^{10}$ 
in length scale. With accurate treatment of atomic and molecular physics,
it allows us to study the chemo-thermal
evolution of primordial gas clouds to densities up to $\rho \sim
2\times 10^{-8} {\rm g}\;{\rm cm}^{-3} (n_{\rm H}\sim 10^{16} {\rm
cm}^{-3})$ without assuming any {\it a priori} equation of state;
a six orders of magnitudes improvement over previous three-dimensional
calculations.
We implement an extensive chemistry network for hydrogen, helium and deuterium.
All the relevant atomic and molecular cooling and heating processes,
including cooling by collision-induced continuum emission, are
implemented.  For calculating optically thick \HH cooling at high
densities, we use the Sobolev method (Sobolev 1960)
and evaluate the molecular line
opacities for a few hundred lines.  We validate the accuracy of the
method by performing a spherical collapse test and comparing the
results with those of accurate one-dimensional calculations that treat
the line radiative transfer problem in a fully self-consistent manner.

We then perform a cosmological simulation adopting the standard
$\Lambda$CDM model.  Dense gas clumps are formed at the centers of low
mass ($\sim 10^{5-6} M_{\odot}$) dark matter halos at redshifts $z\sim
20$, and they collapse gravitationally when the cloud mass exceeds a
few hundred solar masses.  To examine possible gas fragmentation owing to
thermal instability, we compute explicitly the growth rate of isobaric
perturbations. We show that the cloud core does not fragment in
either the low-density ($n_{\rm H} \sim 10^{10} {\rm cm}^{-3}$) or
high-density $(\sim 10^{15} {\rm cm}^{-3}$) regimes, where gas cooling
rate is increased owing to three-body molecule formation and 
collision-induced emission, respectively.  
The cloud core becomes marginally unstable against chemo-thermal instability in the
low-density regime. However, since the core is already compact at that
point and correspondingly the sound-crossing time as well as the
free-fall time are short, or comparable to the perturbation growth
timescale, it does not fragment. Run-away cooling simply leads to fast
condensation of the core to form a single proto-stellar seed. We also
show that the core remains stable against gravitational deformation
and fragmentation throughout the evolution.  We trace in Lagrangian
space the gas elements that end up at the center of the cloud, and
study the evolution of the specific angular momentum. We show that,
during the final dynamical collapse, small angular momentum material
collapses faster than the rest of the gas and selectively sinks
inwards.  Consequently, the central regions have little specific
angular momentum, and rotation does not halt collapse. 
With the large physical dynamic range of our simulation, we,
for the first time, obtain an accurate gas mass accretion rate 
within a 10$M_{\odot}$ innermost region around the protostar.
The protostar is accreting the surrounding hot ($T\sim 2000$K) gas
at a rate of $\dot{M}> 10^{-2}-10^{-3} M_{\odot}/{\rm yr}$.
From these findings we conclude that primordial stars formed in early
cosmological halos are massive.  We carry out proto-stellar evolution
calculations using the obtained accretion rate.  
The resulting mass of the first star when it reaches the zero-age main 
sequece is $M_{\rm ZAMS}\sim 100 M_{\odot}$, 
and less ($\ga 60 M_{\odot}$) for substantially reduced accretion rates.
\end{abstract}


\section{Introduction}

The current standard theory of cosmic structure formation posits that
the present-day clumpy appearance of the Universe developed through
gravitational amplification of matter density fluctuations generated
in its very early history. The energy content of the Universe and the
basic statistics of the initial density field have been determined
with great accuracy from recent observations of the cosmic microwave
background (Spergel et al. 2006), large-scale structure (Tegmark et
al. 2004; Cole et al. 2005), and distant supernovae (Riess et
al. 2004; Astier et al. 2006). It has become possible to make accurate
predictions for the formation and nonlinear growth of large-scale
structure within this framework.

The standard model based on cold dark matter (CDM) predicts that the
mass variance after the epoch of matter-radiation equality 
has progressively larger
amplitudes on smaller mass scales, indicating that objects form
hierarchically, with smaller ones originating first.  Although the
fluctuation amplitudes on scales much smaller than galactic scales are
not known directly from observations, most of the viable candidates
for the dark matter predict nearly scale-invariant fluctuations down to
stellar or even planetary masses.  Such CDM models generically argue
that the first cosmological objects are formed in low mass ($\sim 10^6
M_{\odot}$) dark halos at high redshifts ($z \ga 20$) when primordial
gas condenses via molecular hydrogen cooling (Couchman \& Rees 1986;
Haiman, Thoul \& Loeb 1996; Tegmark et al. 1997; Abel, Bryan \& Norman
2002; Yoshida et al. 2003; see Bromm \& Larson 2004 for a review).
Hierarchical structure formation eventually leads to the emergence of
a population of early generation stars/galaxies which terminate
the Cosmic Dark Ages by emitting the first light (Madau 2000; Cen
2003; Miralda-Escud\'e 2003; Sokasian et al. 2003, 2004; Reed et al. 2005).

The first stars are also believed to be the first source of heavy
element production (those heavier than lithium).  Heavy elements must
have been processed in massive stars and expelled into the
interstellar medium by supernovae, to enable the formation of ordinary
stellar populations.  Early metal-enrichment of the low density
intergalactic medium is also suggested by observations of the
Lyman-$\alpha$ forest (Songaila 2001, 2005; Schaye et al. 2003).
Interestingly, the recent discovery of extremely metal-poor stars
(Christlieb et al. 2002; Frebel et al. 2005) suggests that relics from
very early generation stars exist even today in our own Galaxy.  Stars
with such low metallicities provide valuable information on the
first stars that likely polluted the interstellar medium 
(Iwamoto et al. 2005; Tumlinson 2006).

The study of primordial star formation has a long history.  The
formation of the first cosmological objects via gas condensation by molecular 
hydrogen cooling has been studied for many years (e.g. Saslaw \& Zipoy 1967;
Peebles \& Dicke 1968; Matsuda, Sato, \& Takeda 1969; Kashlinsky \& Rees 1983).
In the context of star formation, the evolution of a collapsing primordial gas
cloud has also been studied extensively (Matsuda, Sato, \& Takeda 1969; Yoneyama 1972; 
Carlberg 1981). 
Palla, Salpeter \& Stahler (1983) used a one-zone model to follow
gas collapse to very high densities.  
They identified a minimum Jeans mass scale of 0.1 solar mass 
from the thermal evolution of the cloud core, suggesting the formation
of low-mass primordial stars.
One-dimensional hydrodynamic simulations of spherical gas collapse were 
performed by a number of researchers (Villere \& Bodenheimer 1987; Omukai \& Nishi 1998; Ripamonti
et al. 2002).  Omukai \& Nishi (1998) included a detailed treatment of
all the relevant chemistry and radiative processes and thus were able
to provide accurate results on the thermal evolution of a collapsing
primordial gas cloud up to stellar densities.  These authors found that,
while the evolution of a spherical primordial gas cloud proceeds in a
roughly self-similar manner, there are a number of differences in the
thermal evolution from that of present-day, metal- and
dust-enriched gas clouds.

Recently, three-dimensional hydrodynamic calculations were performed
by several groups.  Abel et al. (2000; 2002) and Bromm et al. (1999; 2002) studied
the formation and fragmentation of primordial gas clouds in CDM
models.  From various analyses of the simulation results, these
authors conclude that primordial stars (often called ``Population
III'') that form in dark matter matter halos are rather massive.
Yoshida et al. (2003) examined the statistical properties of
primordial star-forming clouds from a large sample of early mini-halos
located in their cosmological simulations.  The chemo-thermal
evolution of the primordial gas clouds is rather complex, being
coupled with the dynamical evolution of the assembly of dark matter
halos (Yoshida et al. 2003; Gao et al. 2005), and simple analytic
models based on crude assumptions almost always fail in predicting the
evolution of primordial gas in CDM halos.

The three-dimensional calculations of Abel et al. (2002; hereafter,
ABN02) were able to follow the gas evolution until a small ($\sim
1M_{\odot}$) fully molecular core formed.  However, since they did not
include the effect of the opacity of the cloud core, the radiative
cooling rate is over-estimated at the cloud center.  Thus their
results for late stage evolution, including the important problem of
fragmentation and mass accretion, are uncertain.  Bromm \& Loeb (2004)
made similar approximations and also continued their simulation by
creating a sink particle to study the late time accretion phase. None
of these simulations reproduce the correct density, temperature, and
velocity profiles in the vicinity the protostar, which are among the
most important quantities for proto-stellar evolution.  A critical
technique we describe in the present paper is computation of molecular
line opacities. With the implementation of optically thick line
cooling, the gas evolution can be followed to much higher densities
than in the previous studies. In principle, evaluating the net line
cooling rate at such high densities involves costly line radiative
transfer.  We circumvent this by using a local Sobolev method which
employs local velocity information as well as knowledge of the density
and temperature.  We show that the method works well in problems of
collapsing gas clouds, in terms of computation of radiative cooling
rates and resulting density and temperature structure.  We apply this
technique to cosmological simulations.

An important, outstanding issue is to determine whether or not a gas
cloud fragments into multiple objects during its collapse.  If it does
fragment, it could yield multiple stars, suggesting the possibility of
star-cluster formation including low-mass stars (Sabano \& Yoshii
1977; Silk 1983).  To the contrary, if the gas cloud remains stable
against fragmentation throughout its evolution, it will lead to the
formation of a single proto-stellar seed, surrounded by a large amount
of infalling gas - a typical condition for a proposed scenario of the
formation of massive stars (Larson \& Starrfield 1971; see the review
by Larson 2003).  We show that the gas cloud formed at the center of a
high-redshift halo remains stable against fragmentation, to form a
single proto-stellar seed.  We also find that the rate of gas
accretion is very high.  The mass accretion rate onto a protostar
largely affects the final stellar mass (Omukai \& Palla 2001; 2003).
Stahler, Palla, \& Salpeter (1986a,b) suggested that large accretion
rates ($> 10^{-3} M_{\odot}\;{\rm yr}^{-1}$) are expected for
primordial protostars because of high gas temperatures in the
infalling surrounding gas.  For a better evaluation of the mass
accretion rate, Omukai \& Nishi (1998) use a self-similar solution to
derive a time-dependent rate, while Tan \& McKee (2004) consider
effects of rotation in accretion dynamics around proto-stellar disks.
We address this issue using fully self-consistent cosmological
simulations.  Finally, we carry out proto-stellar evolution
calculations and determine the resulting stellar mass on the
assumption that the true gas mass accretion rate and its
time-dependence is close to the instantaneous accretion rate at the
protostar formation epoch.

The rest of the paper is organized as follows.  In Section 2, we
describe the chemistry network used for our simulations. The relevant
cooling and heating processes are described in Section 3. There, we
also provide details of numerical implementation. In Section 4, we
present the results of a test problem of a spherically collapsing gas.
We compare our results to previous work using one-dimensional
codes. We present the results of our $\Lambda$CDM cosmological
simulations in Section 5.  Implications of the main conclusions are
discussed in Section 6.

\section{Primordial gas chemistry}

The reaction network for hydrogen and helium (e$^{-}$, H, H$^+$, He,
He$^{+}$, He$^{++}$, H$_{2}$, H$_{2}^{+}$, H$^{-}$) is largely based
on that of Abel et al. (1997) and Galli \& Palla (1998; hereafter,
GP98).  The rate coefficients are summarized in Table 1.  In this
section, we describe updates in detail.

\subsection{Molecule formation}
In a low density primordial gas, 
the main formation path for hydrogen molecules is
via the H$^-$ channel:
 
\begin{equation}
{\rm H} + {\rm e} \rightarrow {\rm H}^-  + h\nu \;\;\;\;\;\;({\rm reaction\;\;7}), 
\end{equation}
\begin{equation}
{\rm H} + {\rm H}^- \rightarrow {\rm H}_2 + e \;\;\;\;\;\;({\rm reaction\;\;8}), 
\end{equation}
where electrons are used as catalysts. \HH formation via the H$_2^+$ channel
(reaction 9, 10) is dominant at very high redshifts $z>200$ where cosmic 
microwave background photons destroy H$^-$ by photo-detachment
(see also a recent calculation of Hirata \& Padmanabhan 2006, 
which shows the importance of reactions involving HeH$+$ ions 
in \HH production at such high redshifts.)
There is considerable variation among published experimental
data and theoretical calculations for the reaction rate $k_8$.
However, uncertainty in this rate 
will have little effect on the \HH abundance, because the reaction
is always much faster than the formation of H$^-$.
We use the fit by GP98 to the calculations
of Launay et al. (1991) for $k_8$. 
The rate coefficients for mutual neutralization
\begin{equation}
{\rm H}^- + {\rm H}^+ \rightarrow {\rm H}_2\;\;\;\;\;\;({\rm reaction\;\; 18}), 
\end{equation}
are also uncertain, particularly at low temperatures (Glover et al. 2006).
This reaction affects the \HH abundance significantly only when 
the ionization fraction is greater than $\sim 0.01$ (Palla \& Zinnecker 1987).
In the calculations presented below, the ionization fraction 
is always smaller than $10^{-3}$, and hence the uncertainty 
in the reaction rate is unimportant. 
We use the fit given in GP98. We note that a better determination of
this rate is clearly needed for calculations of, for example, 
the evolution of cooling gas in relic \HII regions (Oh \& Haiman 2002; 
Glover et al. 2006; Yoshida 2006; Susa \& Umemura 2006).

\subsection{Three-body reactions}
At high densities ($n_{\rm H} > 10^8 {\rm cm}^{-3}$), 
hydrogen molecules are formed by the following rapid 
reactions
\begin{equation}
{\rm H} + {\rm H} + {\rm H}  \rightarrow {\rm H}_2  + {\rm H}\;\;\;\;\;\;({\rm reaction\;\; 22}) , 
\end{equation}
and
\begin{equation}
{\rm H} + {\rm H} + {\rm H}_2 \rightarrow 2{\rm H}_2\;\;\;\;\;\;\;\;({\rm reaction\;\; 23}). 
\end{equation}
We use the reaction coefficients as given in Palla, Salpeter, \& 
Stahler (1983):
\begin{equation}
k_{22} = 5.5\times 10^{-29}\; T^{-1} \;\;{\rm cm}^6 s^{-1},
\end{equation}
and
\begin{equation}
k_{23} = 6.875\times 10^{-30}\; T^{-1} \;\;{\rm cm}^6 s^{-1}.
\end{equation}
\noindent Note that the net formation rate scales with the cube of density.
The three-body reactions are very efficient in converting hydrogen atoms
to molecules at densities $n > 10^8 {\rm cm}^{-3}$.

Krstic, Janev \& Shultz (2003) recently revised reaction rates for
three-body reactions involving protons,
\begin{equation}
{\rm H} + {\rm H} + {\rm H}^+ \rightarrow {\rm H}_2  + {\rm H}^+,  
\end{equation}
\begin{equation}
{\rm H} + {\rm H} + {\rm H}^+ \rightarrow {\rm H}_2^+  + {\rm H}, 
\end{equation}
and found the rates are as large as those for reactions (22), (23).
Since the ionization fraction is extremely small in the primordial gas
at high densities and low temperatures ($T<3000$ K), which
is the range of our interest, these reactions are unimportant 
in the calculations presented in what follows.
We note, however, that when the ionization fraction (hence proton
abundance) is large in the proto-stellar gas cloud
in the final adiabatic phase, or under strong ultra-violet radiation, 
inclusion of the reactions could be important.

\subsection{Collisional dissociation of \HH}
\label{sec:diss}
In a neutral primordial gas, \HH molecules
are destroyed mostly by collisions with H atoms
and \HH molecules:
\begin{equation}
{\rm H}_2  + {\rm H}  \rightarrow 3{\rm H}\;\;\;\;\;\;\;\;({\rm reaction\;\; 11}),
\label{eq:dissociation}
\end{equation}
\begin{equation}
{\rm H}_2  + {\rm H}_2  \rightarrow {\rm H} + {\rm H} + {\rm H}_2 \;\;\;\;\;\;\;\;({\rm reaction\;\; 25})
\end{equation}
We use the collision cross-sections of Martin et al. (1996)
for reaction 11 that include dissociation from excited 
levels at high densities.
For dissociation by collisions with \HH (reaction 25),
we use the high density limit of  Palla et al. (1983)
\begin{equation}
k_{25} = 8.125\times 10^{-8}\;T^{-1/2}
\exp\left(-\frac{52000}{T}\right) \times \left[1.0-\exp\left(-\frac{6000}{T}\right)\right].
\end{equation}

Dissociation by collisions with helium atoms
\begin{equation}
 {\rm H}_2  + {\rm He}  \rightarrow  {\rm H} + {\rm H} + {\rm He}
\end{equation}
and its reverse reaction (three-body reaction involving helium) have much 
smaller reaction rates than those involving
collisions with hydrogen (Dove et al. 1987), and thus we ignore these processes.

We include the charge exchange reaction between H and He
(reaction 14, 15, see Glover \& Brand 2003) and update the reaction rate for 
${\rm H}+{\rm H}_2^+ \rightarrow {\rm H}^++{\rm H}_2$
(reaction 12)
with the new fit given by Savin et al. (2004a,b).
These processes become important in calculations of, for example, the evolution of relic
\HII regions (Yoshida 2006), but do not affect the 
calculations in the present paper.


\subsection{The adiabatic index}
We write the equation of state for a gas consisting of atoms 
and molecules as
\begin{equation}
P = \frac{k_{\rm B}T}{\mu m_{\rm p}}\rho = (\gamma - 1)\rho u,
\end{equation}
where the adiabatic index is computed from
\begin{equation}
\gamma - 1 = \frac{1}{\mu}\left( \sum_{i} \frac{1}{(\gamma_i - 1)}\right)^{-1}.
\end{equation}
The summation is taken over all chemical species including helium.
For the adiabatic exponent for \HH, we use the formula in Landau \& Lifshitz (1984),
\begin{equation}
\frac{1}{\gamma_{\rm H_2} - 1} =  \frac{1}{2}\left[5+2x^2\frac{e^x}{(e^x-1)^2}\right],
\end{equation}
where the last term with $x=6100 {\rm K}/T$ accounts for vibrational
degrees of freedom of hydrogen molecules. At very high pressures 
($P \gg 10^4 {\rm dyne}\;{\rm cm}^{-2}$), 
various non-ideal gas effects become important and the equation of state needs 
to be modified (Saumon, Chabrier, \& van Horn 1995; Ripamonti et al. 2002),
but this is outside the range of gas phases we consider here.

\section{Cooling and heating rates}
Radiative cooling processes owing to excitation, ionization, and recombination
of hydrogen and helium atoms are well established. 
We refer the readers to Cen (1992)\footnote{The recombination rate for He$^{++}$ in
Cen (1992) has an incorrect temperature dependence. See 
Fukugita \& Kawasaki (1994) for the correct scaling.}
, Fukugita \& Kawasaki (1994), and Abel et al. (1997).
We note that these cooling processes are generally unimportant
in studies of the evolution of a neutral primordial gas.
Below we describe more relevant molecular cooling and 
chemical heating processes in detail.

\subsection{Cooling by H$_{2}$ and HD molecules}
We use the cooling rate of Galli \& Palla (1998) for H$_{2}$
line cooling in the low density limit:
\begin{equation}
\Lambda_{\rm H_2}(n\rightarrow 0)={\rm dex}[-103.0+97.59 \log T-48.05 (\log T)^2
+10.80(\log T)^3-0.9032 (\log T)^4].
\end{equation}
At high densities, formation of \HH molecules takes place
and molecular fraction increases rapidly, making \HH line cooling
a dominant cooling process.
At high densities, all energy levels are populated according 
to local thermodynamic equilibrium (LTE). 
An accurate fit for the optically thin cooling rate 
is given by Hollenbach \& McKee (1979):
\begin{equation}
\Lambda_{\rm LTE} = \Lambda_{\rm rot} + \Lambda_{\rm vib},
\label{eq:HM79cool}
\end{equation}
where
\begin{equation}
\Lambda_{\rm rot} = \frac{9.5\times 10^{-22}\;T_3^{3.76}}{1+0.12 \;T_3^{2.1}} 
\exp\left[-\left(\frac{0.13}{T_3}\right)^3\right]+ 3\times 10^{-24}\exp\left[-\left(\frac{0.51}{T_3}\right)\right],
\label{eq:HM79cool_rot}
\end{equation}
and
\begin{equation}
\Lambda_{\rm vib} = 6.7\times 10^{-19}\exp\left[-\left(\frac{5.86}{T_3}\right)\right] 
+ 1.6\times 10^{-18}\exp\left[-\left(\frac{11.7}{T_3}\right)\right],\;\;\; T_3=\frac{T}{1000 {\rm K}} \, .
\label{eq:HM79cool_vib}
\end{equation}

\noindent The low density limit and the high density limit are bridged as
\begin{equation}
\Lambda_{\rm H_2} = \frac{\Lambda_{\rm H_2}({\rm LTE})}
{1+n_{\rm cr}/n_{\rm H}} \, ,
\end{equation}
where
\begin{equation}
\frac{n_{\rm cr}}{n_{\rm H}}=\frac{\Lambda_{\rm H_2}({\rm LTE})}
{\Lambda_{\rm H_2}(n\rightarrow 0)} \, .
\end{equation}

For HD cooling, we use the formulation of Flower et al. (2000)
\footnote{The HD cooling function of Flower et al.
does not include transitions between high vibrational levels but
the contribution to the total cooling rate at the relevant densities 
are unimportant, as shown by Lipovka et al. (2005).}.
We show the cooling rates in Fig. \ref{fig:H2line}.
There, we assume a fractional abundance of [HD/H$_2$] = $10^{-3}$.
Cooling by HD molecules is important only at low temperatures ($T < 200 {\rm K}$)
and low densities ($n_{\rm H} < 10^8$); otherwise
\HH cooling dominates. 
 We have carried out a spherical collapse simulation 
(see Section \ref{sec:sphere}) with full deuterium chemistry 
for five species (D, D$^+$, D$^-$, HD, HD$^+$),
using the extensive reaction network of Nakamura \& Umemura (2002).
We find that the fractional ratio of [HD/H$_2$] 
has a maximum of $\sim 10^{-3}$ at $n_{\rm H} \sim 10^4\;{\rm cm}^{-3}$,
where the gas temperature is about 200 K
(see Fig. \ref{fig:d_T} and also Omukai et al. 2005). 
Therefore, HD cooling is unimportant for the thermal evolution
of initially neutral primordial gas clouds.
The role of HD cooling in partially ionized gas will be studied
elsewhere (Yoshida et al. 2006, in preparation).

\begin{inlinefigure}
\resizebox{10cm}{!}{\includegraphics{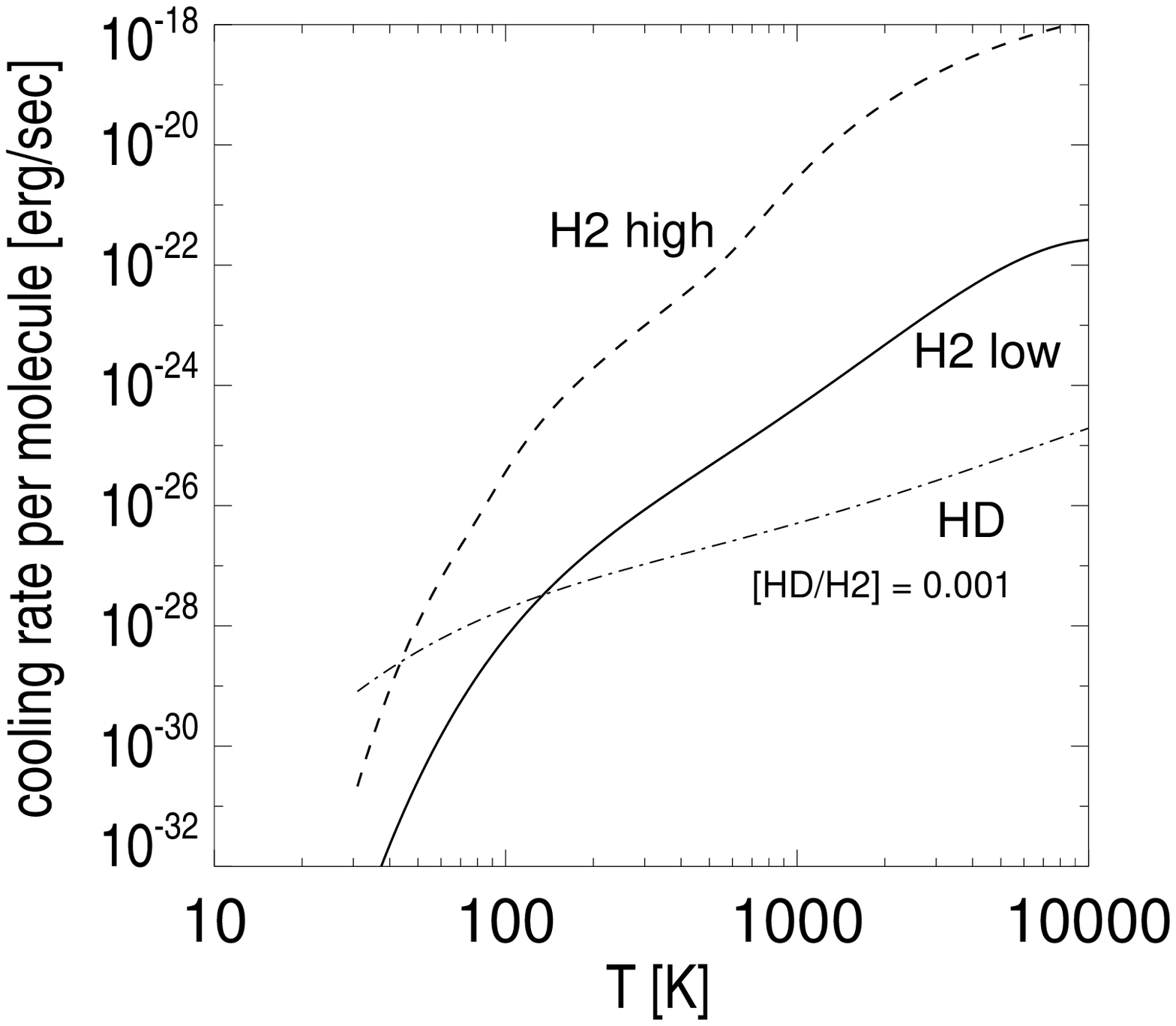}}
\caption{Radiative cooling rates per molecule as a function of temperature.
We show the low-density limit for H$_{2}$ (solid line) and HD (dot-dashed line)
assuming $n_{\rm H}=1 \; {\rm cm}^{-3}$ and the 
relative abundance HD/H$_{2}$ = 0.001. 
The dashed line is the high-density limit of the H$_{2}$
cooling function.
\label{fig:H2line}}
\end{inlinefigure}

\subsection{Chemical cooling and heating}
We include the heat gain associated with the formation of
hydrogen molecules via three-body reactions as
\begin{equation}
\Gamma_{\rm H2, 3b} = \epsilon_{{\rm H}_2} \frac{{\rm d} n_{{\rm H}_2}}{{\rm d}t} \, ,
\end{equation}
where $\epsilon_{{\rm H}_2} =  4.48 {\rm eV}$ is the molecular binding energy.
We also account for heat loss owing to collisional dissociation (Section \ref{sec:diss})
in the same manner.
The two-body formation process of \HH molecules, either via H$^-$ (reaction 8) 
or H$_2^+$ (reaction 10),
is exothermic, but deposits its energy into \HH through
rotational and vibrational excitation.
These reactions can be important sources of heat at high densities 
where collisional de-excitation removes the excitation energy. Following
Hollenbach \& McKee (1979) and Shapiro \& Kang (1987), we calculate
the heating rates as
\begin{equation}
\Gamma({\rm H}^-) = k_{8} n_{\Hminus} n_{\rm H} [3.53 (1 + n_{\rm cr}/n_{\rm H})^{-1}] 
\;\;{\rm eV}{\rm cm}^{-3} {\rm s}^{-1},
\end{equation}
\begin{equation}
\Gamma({\rm H}_2^+) = k_{10} n_{\Htwoplus} n_{\rm H} [1.83 (1 + n_{\rm cr}/n_{\rm H})^{-1}] 
\;\;{\rm eV}{\rm cm}^{-3} {\rm s}^{-1},
\end{equation}
where the critical density is given by
\begin{equation}
n_{\rm cr} = 10^6 T^{-1/2} 
\left[ 1.6 y_{\rm H} \exp[-(400/T)^2] 
+ 1.4\; y_{\Htwo} \exp \left(-\frac{12000}{T+1200}\right) \right]^{-1}
{\rm cm}^{-3},
\end{equation}
and $y_{\rm H}$ and $y_{\Htwo}$ are the number fractions
of hydrogen atoms and hydrogen molecules, respectively.

\subsection{Optically thick \HH line cooling}
\label{sec:sobolev}
When the gas density and the molecular fraction are
high, the cloud becomes opaque to molecular lines
and then \HH line cooling becomes inefficient.
The net cooling rate can be expressed as
\begin{equation}
\Lambda_{{\rm H}_2, {\rm thick}}=\sum_{u,l} h\nu_{ul}\;\beta_{{\rm esc}, ul}\; A_{ul}\; n_{u} \, ,
\label{eq:H2thick}
\end{equation}
where $n_{u}$ is the population density of hydrogen molecules in the upper energy level $u$,
$A_{ul}$ is the Einstein coefficient for spontaneous transition, 
$\beta_{{\rm esc}, ul}$ is the probability for an emitted line photon to escape without absorption,
and $h\nu_{ul} = \Delta E_{ul}$ is the energy difference between the two levels.
In the relevant temperature range $T \sim 1000-2000$K,
we consider rotational levels from $J=0$ to 20, and vibrational levels $v=0,1,2$.
We use the radiative transition rates $A_{ul}$ for these levels from Turner, Kirby-Docken, Dalgarno (1977).
We calculate the
vibrational energies $E(v,J)$ following Borysow, Frommhold \& Moraldi (1989).
Note that, although the net cooling rate can be directly obtained from 
equation (\ref{eq:H2thick}), we use it to evaluate the reduction factor
$f_{\rm red} = \Lambda_{\rm thick}/\Lambda_{\rm thin}$. For computational efficiency,
we compute the cooling rate as a function of temperature using equations (\ref{eq:HM79cool})-(\ref{eq:HM79cool_vib}),
and multiply the rate by $f_{\rm red}$ when $n_{\rm H} > 10^{9} {\rm cm}^{-3}$. 

In order to calculate the escape probability, we first evaluate the opacity for each molecular 
line as follows.
The absorption coefficient for a transition from
$l$-level to $u$-level is 
\begin{equation}
\alpha_{lu}=\frac{\Delta E_{lu}}{4\pi} \;n_{l}B_{lu} 
\left[1-\exp\left(\frac{-\Delta E_{lu}}{kT}\right)\right] \phi(\nu) \, ,
\end{equation}
where $n_{l}$ is the number density of hydrogen molecules in level $l$, $B_{lu}$ is Einstein's B-coefficient, and $\phi(\nu)$ is
the line profile. We approximate the line profile by a Gaussian function
\begin{equation}
\phi(\nu) = \frac{1}{\sqrt{\pi}\Delta\nu_{D}} 
\exp\left[-\frac{(\nu-\nu_0)^2}{\Delta \nu_{\rm D}^2}\right],
\end{equation}
with the thermal Doppler width $\Delta\nu_{D} = (\nu_0/c)/\sqrt{kT/m_{\rm H}}$. 
We then compute the opacity at the line center as
\begin{equation}
\tau_{lu}=\alpha_{lu} L,
\end{equation}
where $L$ is the characteristic length scale,
which is approximately the cloud core size.
Since the absorption coefficients are computed in a straightforward manner,
although somewhat costly, 
the remaining key task is the evaluation of the length scale $L$.

It is important to recall that calculation of the photon
escape probability is essentially a line transfer problem,
for which, in principle, demanding radiative transfer calculations are
needed to obtain accurate results.
However, by noting that the important quantity we need 
is the effective gas cooling rate, we can formulate
a reasonable and 
well-motivated approximation.
To this end, we decided to use the escape probability method
that is widely used
in the study of stellar winds and planetary nebulae 
(Castor 1970; Goldreich \& Kwan 1974).
Consider a photon traveling in a gas in one direction 
along which a constant velocity gradient exists. 
The escape of photons is greatly enhanced by the presence of 
macroscopic velocity fields, i.e. large velocity gradients;
when an emitted photon has traveled one Sobolev length
to the point where the profile is Doppler-shifted
by one characteristic width (line width), it can travel
unimpeded and will escape from the local neighborhood.

We calculate the Sobolev length along a line-of-sight
as
\begin{equation}
L_{r} = \frac{\;v_{\rm thermal}\;}{|dV_{r}/dr|},
\end{equation}
where $v_{\rm thermal} = \sqrt{kT/m_{\rm H}}$ is the thermal
velocity of \HH molecules,
and $V_{r}$ is the fluid velocity in the direction.
A suitable angle-average must be computed
in order to obtain the net escape probability.
For a spherical cloud, the escape probability 
is given by
\begin{equation}
\beta_{\rm esc} = \frac{1-\exp(-\tau)}{\tau} \, ,
\label{eq:escapeprob}
\end{equation}
with $\tau = \alpha L_{r}$
(Castor 1970; de Jong, Dalgarno \& Chu 1975).
In three-dimensional calculations, in which we do 
not know the exact geometry and alignment of an object,
we compute the Sobolev 
length and the escape probability
in three arbitrary orthogonal directions 
and take the mean as
\begin{equation}
\beta = \frac{\beta_x + \beta_y + \beta_z}{3} \, .
\label{eq:esc_ave}
\end{equation}

We first test the implementation in this section,
and the overall accuracy of the method is examined in Section \ref{sec:sphere}.
Consider a homologous contracting sphere with size $R=0.01$ pc. 
The velocity at radius $r$ is given by 
\begin{equation}
v(r)= r V/R \, ,
\end{equation}
where $V=10$ km/sec is the radial (inward) velocity at R. We set the gas temperature to 
be a constant at $T=1000$K and assume the gas is fully molecular with
$n_{\rm H2} = 5.0\times 10^{10} {\rm cm}^{-3}$.
In this case, the optical depth for an emitted photon
(owing to a level transition $u\rightarrow l$) is a constant
\begin{equation}
\tau_{lu}=\frac{\Delta E_{lu}}{4\pi} \;n_{l}B_{lu} 
\left[1-\exp\left(\frac{-\Delta E_{lu}}{kT}\right)\right] \frac{R}{V},
\label{eq:tau64}
\end{equation}
and the escape probability is
\begin{equation}
\beta_{{\rm esc}, lu} = \frac{1-\exp(-\tau_{lu})}{\tau_{lu}}.
\end{equation}
We run a three-dimensional simulation with the above set-up using a half million gas particles
and compute the escape probability (equation [\ref{eq:esc_ave}]) 
at random points in the simulation region.
For illustrative purpose, we consider only the rotational transition
$J = 6\rightarrow 4 (v=0)$, which is one of the strongest lines at $T\sim 1000$K.
Putting all the numerical values into equation (\ref{eq:tau64}),
we find $\tau_{46} = 3.61$ and $\beta_{46}=0.269$.  From the test simulation, 
we obtained a mean value 
of $\bar{\beta}=0.2692$ within $r < 0.75 R$ and a maximum
deviation of less than 3 \%. We did not include the outer parts
of the sphere ($r > 0.75 R$), where the density estimate is affected
by the boundary. These results are quite satisfactory,
and assure us that the numerical implementation is done correctly.
We discuss the overall accuracy of this method in Section \ref{sec:sphere} 
where we perform a gas collapse simulation with density and temperature gradients.
We emphasize that simple local estimates, for example, 
using only local densities and temperatures, do not provide 
correct values for the optically-thick line cooling rate, because 
the escape probability of line photons 
is greatly affected by local velocity gradients
which are highly variable in both space and time.

The Sobolev method can also be extended to calculation of the self-shielding factor 
against photo-dissociation by Lyman-Werner photons. 
The feedback effects of far ultra-violet radiation is often
modeled either by assuming the gas is optically thin to the photo-dissociating 
photons (Machacek, Bryan \& Abel 2001; Ricotti et al. 2002),
or by employing the formula of Draine \& Bertoldi (1996) in cases with simple geometry 
(Omukai 2001).
While accurate results can be obtained by solving radiative transfer
for a number of lines, as done by, e.g., Glover \& Brand (2001) in one-dimension,
it will be useful to devise an approximation.
The first such attempts of accounting for self-shielding have been already made, 
although in a much simplified manner, by Yoshida et al. (2003) and Glover et al. (2006).
Our method (or its simple extension) may provide better solutions than 
these previous works.

\subsection{Cooling by collision-induced emission}
\label{sec:CIE}
At densities greater than $n_{\rm H} \sim 10^{14} {\rm cm}^{-3}$,
hydrogen molecules collide frequently and 
collision pairs can generate an induced electric dipole,
through which either molecule make an energy transition
by emitting a photon. This process is known as
collision-induced emission (CIE), the opposite
process to collision-induced absorption:
\begin{equation}
{\rm H}_{2}(v,J)  + {\rm H}_2  \rightarrow {\rm H}_{2}(v',J')  + {\rm H}_2 + h\nu,
\end{equation}
\begin{equation}
{\rm H}_{2}(v,J)  + {\rm He}  \rightarrow  {\rm H}_{2}(v',J')  + {\rm He} + h\nu.
\end{equation}

\begin{inlinefigure}
\resizebox{10cm}{!}{\includegraphics{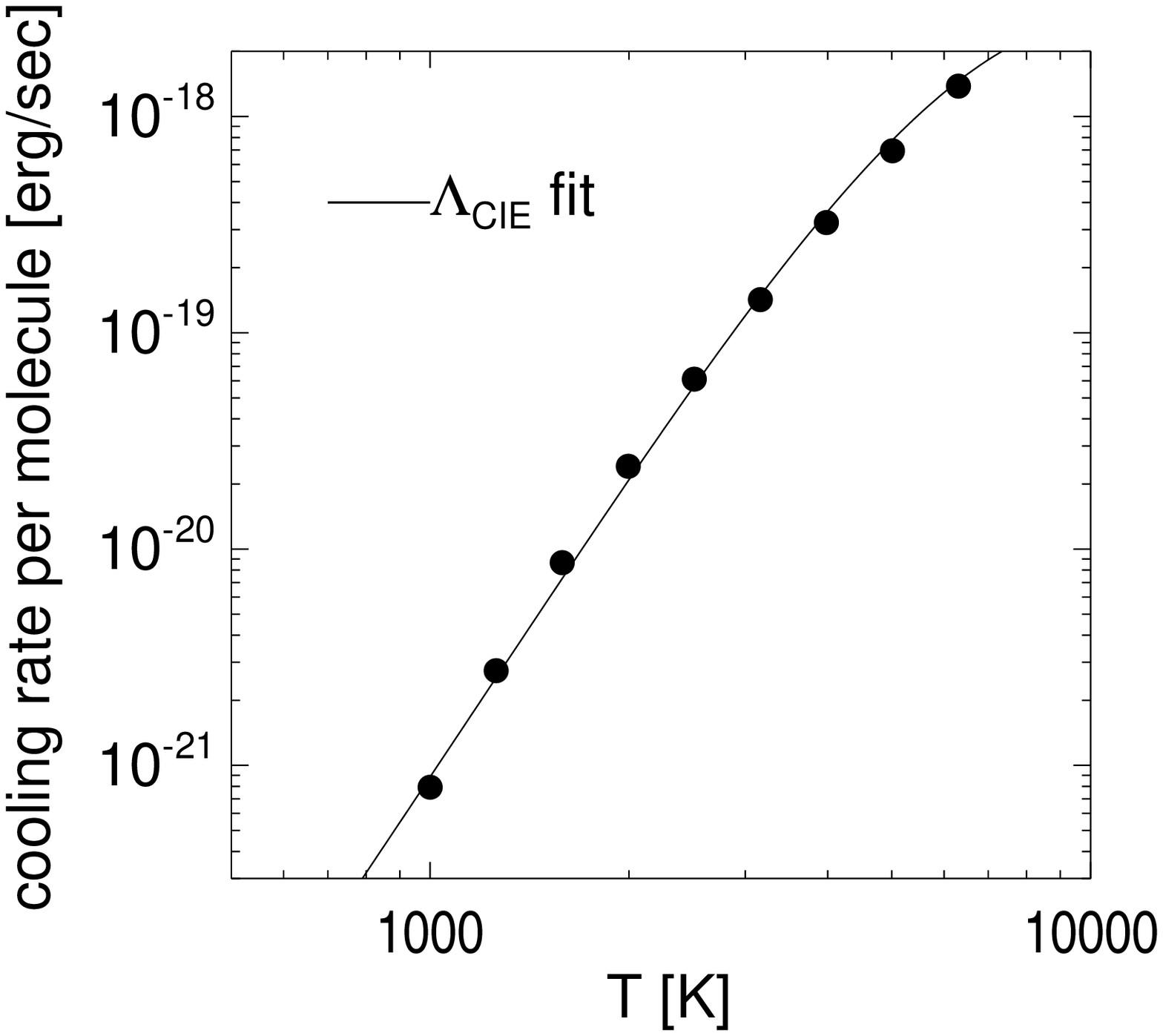}}
\caption{Cooling rate per molecule owing to collision-induced emission
as a function of temperature. 
The solid points are calculated from equation (\ref{eq:eta_CIE}).
We consider only the dominant ${\rm H}_2$-${\rm H}_2$ and ${\rm H}_2$-He collisions
assuming hydrogen is all in molecules.
The solid line is our fit equation (\ref{eq:CIEfit}).
We assume $n_{\rm H2}=10^{15} {\rm cm}^{-3}$ in this plot. 
\label{fig:CIE_rate}}
\end{inlinefigure}

\noindent This process yields very complex spectra, and have an essentially continuum 
appearance.
For \HH-\HH collisions, only fundamental vibrational bands can be noticed
as smooth bumps in the spectra (see Formmhold 1994 for full
account of this process). 
Thus we need to calculate the total emissivity by  integrating the contribution from
each transition:
\begin{equation}
\eta_{\rm CIE} = \frac{2h\nu^3}{c^2}\;\sigma_{\rm CIE}\; n({\rm H}_2)
\exp\left(-\frac{h\nu}{kT}\right) \, .
\label{eq:eta_CIE}
\end{equation}
We use the cross-sections of Jorgensen et al. (2000)
and Borysow, Jorgensen, \& Fu (2001) and Borysow (2002).
In practice, for simulations of a collapsing primordial gas,
CIE cooling is important in the phase
where the gas is nearly fully molecular (Omukai 2001). We consider \HH-\HH
and \HH-He collisions, where the latter gives a minor contribution
to the cooling rate. In practice, we use a simple
fit for the total cooling rate as a function of temperature as
\begin{eqnarray} 
\Lambda_{\rm CIE} &=& {\rm dex}[-116.6 + 96.34\times \log T  \nonumber \\ 
&-& 47.153\times (\log T)^2 
+ 10.744\times (\log T)^3 - 0.916\times (\log T)^4].
\label{eq:CIEfit}
\end{eqnarray}
In Fig. \ref{fig:CIE_rate}, we compare the fit with the accurate cooling rate
obtained directly from the cross-sections. Clearly, equation (\ref{eq:CIEfit})
provides an excellent fit over the plotted temperature range. 

\subsection{Limitation of the simulations}
We successfully implemented the relevant atomic and molecular physics
to follow the gas evolution to very high densities, $n_{\rm H} \sim 10^{16} {\rm cm}^{-3}$.
What we have included is sufficient for the present work, but in
principle
we will need to implement additional physics to probe 
higher densities.
We here briefly describe necessary modifications in order to follow the evolution 
of the proto-stellar gas further, up to stellar densities.
First, for densities much larger than $10^{15} {\rm cm}^{-3}$,
the reaction time scale becomes significantly shorter than
the dynamical time. Then, we can use equilibrium chemistry which would actually 
{\it simplify} our handling of chemistry evolution. The species abundances
can be determined from the Saha-Boltzmann equations, for example.
Second, for $n_{\rm H} > 10^{16} {\rm cm}^{-3}$, the gas cloud becomes opaque 
even to continuum emission, and
then we need to evaluate continuum opacity.
We have already implemented the calculation of the continuum opacity
using the table of the Planck opacity for primordial gases (Lenzuni, Chernoff, 
\& Salpeter 1991; Mayer \& Duschl 2005).
Third, when dissociation of hydrogen molecules
is completed at $T \sim 5000$ K, 
there will be no further mechanisms that enable the gas
to lose its thermal energy, and then the gas temperature
increases following the so-called 
adiabatic track. The equation of state in the high pressure
regime must be modified
to account for non-ideal gas effects (Saumon et al. 1995; Ripamonti et al. 2002).
In future work, we pursue simulating the formation of protostars
by employing much higher mass resolution and all these
necessary physics.

\subsection{Time integration}
\label{sec:time}
We incorporate the chemistry solver in the 
parallel $N$-body/Smoothed Particle Hydrodynamics (SPH) 
code GADGET-2 (Springel 2005)
in the following manner. The code computes 
gravitational and hydrodynamic forces and updates 
the particles positions and velocities. All the force
terms as well as other quantities are evaluated in double
precision in our simulations. For gas particles,
the code also updates the density and specific entropy.
In particular, we employ the fully conservative 
formulation of SPH (Springel \& Hernquist 2002), which
maintains energy and entropy conservation even when
smoothing lengths vary adaptively (Hernquist 1993).
This refinement to the SPH method is crucial for
accurately describing the evolution of gas over the
large dynamic range required to follow primordial
star formation in a cosmological context.
At the end of one time step, we evolve the fractions of
the chemical species, and add an extra heating (cooling)
term which arises from chemical reactions. 

We use a backward difference formula (Anninos et al. 1997) 
\begin{equation}
n^{t+\Delta t} = \frac{C^{t+\Delta t} \Delta t+ n^t}{1+D^{t+\Delta t} \Delta t} \, .
\end{equation}

\noindent For each gas particle, we supplement the usual Courant condition with two additional
constraints so that the time step does not exceed the characteristic cooling time
and the characteristic chemical reaction time.
We monitor the cooling time 
\begin{equation}
\Delta t_{\rm cool} = e_{\rm tol} \frac{\;T\;}{\;\dot{T}\;}
\end{equation}
and use the rate of change of the electron number density and that
of neutral hydrogen to measure the characteristic chemical reaction time
\begin{equation}
\Delta t_{\rm chem} = e_{\rm tol} \;
\min\{ \frac{\;n_{\rm e}\;}{\;\dot{n}_{\rm e}\;}\;,\;
\frac{\;n_{\rm HI}\;}{\;\dot{n}_{\rm HI}\;} \}.
\end{equation}
Setting $e_{\rm tol}=0.1$ suffices for following the evolution
of low density gas ($ n < 10^7 {\rm cm}^{-3}$). At high densities, 
particularly at $ n \sim 10^8-10^{11} {\rm cm}^{-3}$, 
rapid reactions release a significant amount of heat.
After some experiments, we found that the chemistry part becomes numerically unstable
with $e_{\rm tol}=0.1$ and concluded that setting $e_{\rm tol}=0.01$
allows stable time integration at $ n > 10^7 {\rm cm}^{-3}$.

\section{Spherical collapse test}
\label{sec:sphere}
In this section, we test our numerical techniques using a spherical collapse problem.
We follow the evolution of primordial gas in a dark matter halo by setting 
up a gas sphere embedded in a NFW (Navarro, Frenk, White 1997) potential
\begin{equation}
\rho (r) = \frac{\rho_{\rm s}}{(r/r_{\rm s}) (1+(r/r_{\rm s})^2)},
\end{equation}
where $r_{\rm s}, \rho_{\rm s}$ are scale radius and density, respectively. 
The initial gas density is set to be an isothermal $\beta$ profile
\begin{equation}
\rho_{\rm g} (r) = \frac{\rho_{\rm g,0}}{[1+(r/r_{\rm s})^2]^{3\beta/2}}.
\label{eq:iso}
\end{equation}
We are interested in the gas evolution after the 
gas cloud becomes self-gravitating, and so details of the initial
density profile do not matter. For simplicity, we set $\beta = 1$.
The halo mass is set to be $5\times 10^5 M_{\odot}$ and we assume the
baryon fraction to be 0.05. We distribute 4 million particles 
according to equation (\ref{eq:iso}) and evolve the system.

Fig. \ref{fig:d_T} shows the distribution of gas in 
a temperature-density phase plane when the central density is
$5\times 10^{15} {\rm cm}^{-3}$. In the figure, characteristic
features are marked as regions A-G.
The bottom panel in Fig. \ref{fig:d_T} shows the corresponding 
molecular fraction distribution. All the features are explained
as closely related to the thermal evolution. See the caption
for a brief explanation.
The overall evolution of the central gas cloud after it
undergoes a run-away collapse is consistent
with the spherically symmetric calculation of Omukai \& Nishi (1998;
hereafter, ON98).
We have checked the radial profiles of density, temperature, velocity
and molecular fraction. These quantities are quite similar
to the late time evolution of the ON98 calculation
until the central gas density reaches $\sim 10^{16} {\rm cm}^{-3}$
(up to the fourth output in Fig. 1 of ON98).

\begin{inlinefigure}
\resizebox{16cm}{!}{\includegraphics{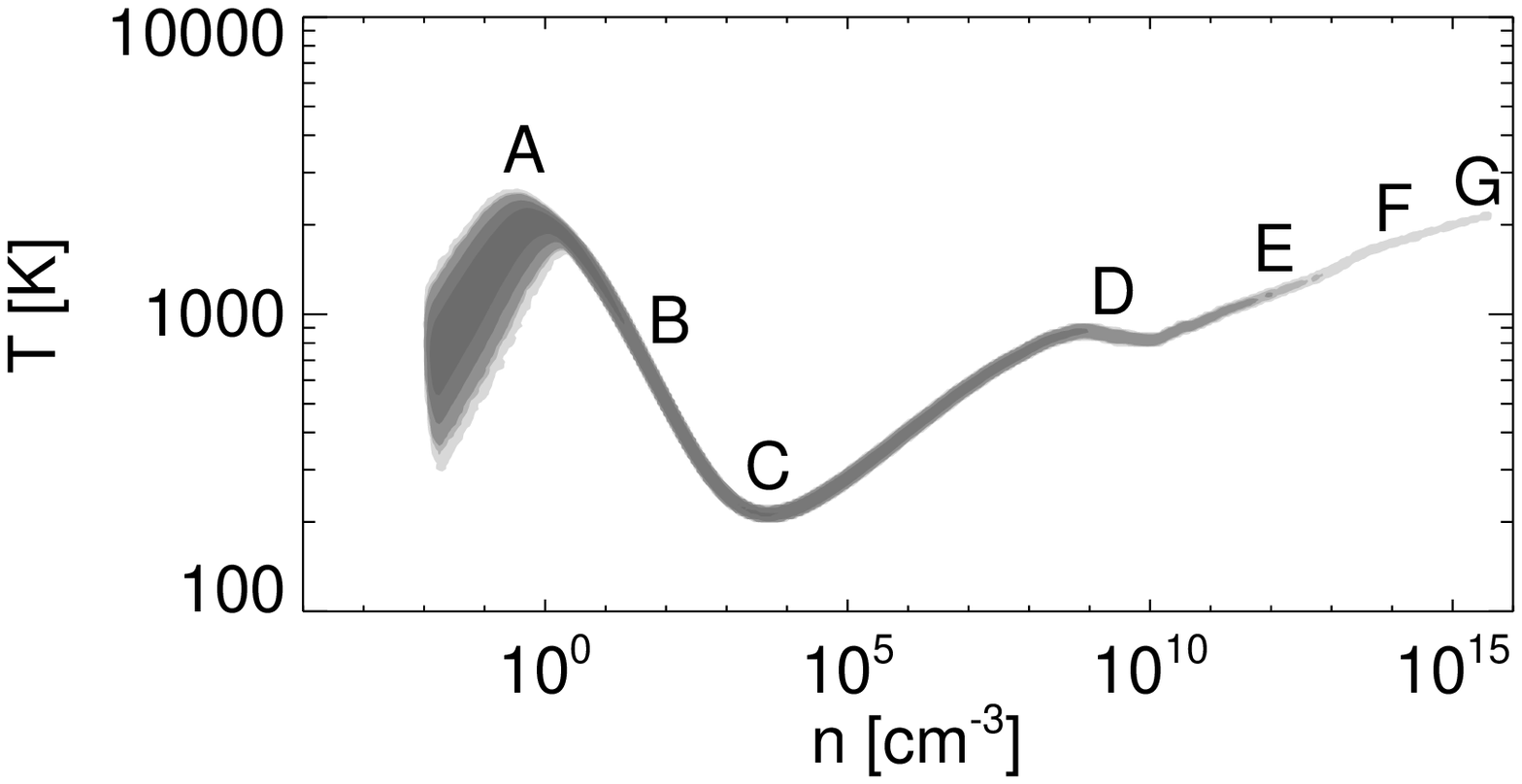}}
\resizebox{16cm}{!}{\includegraphics{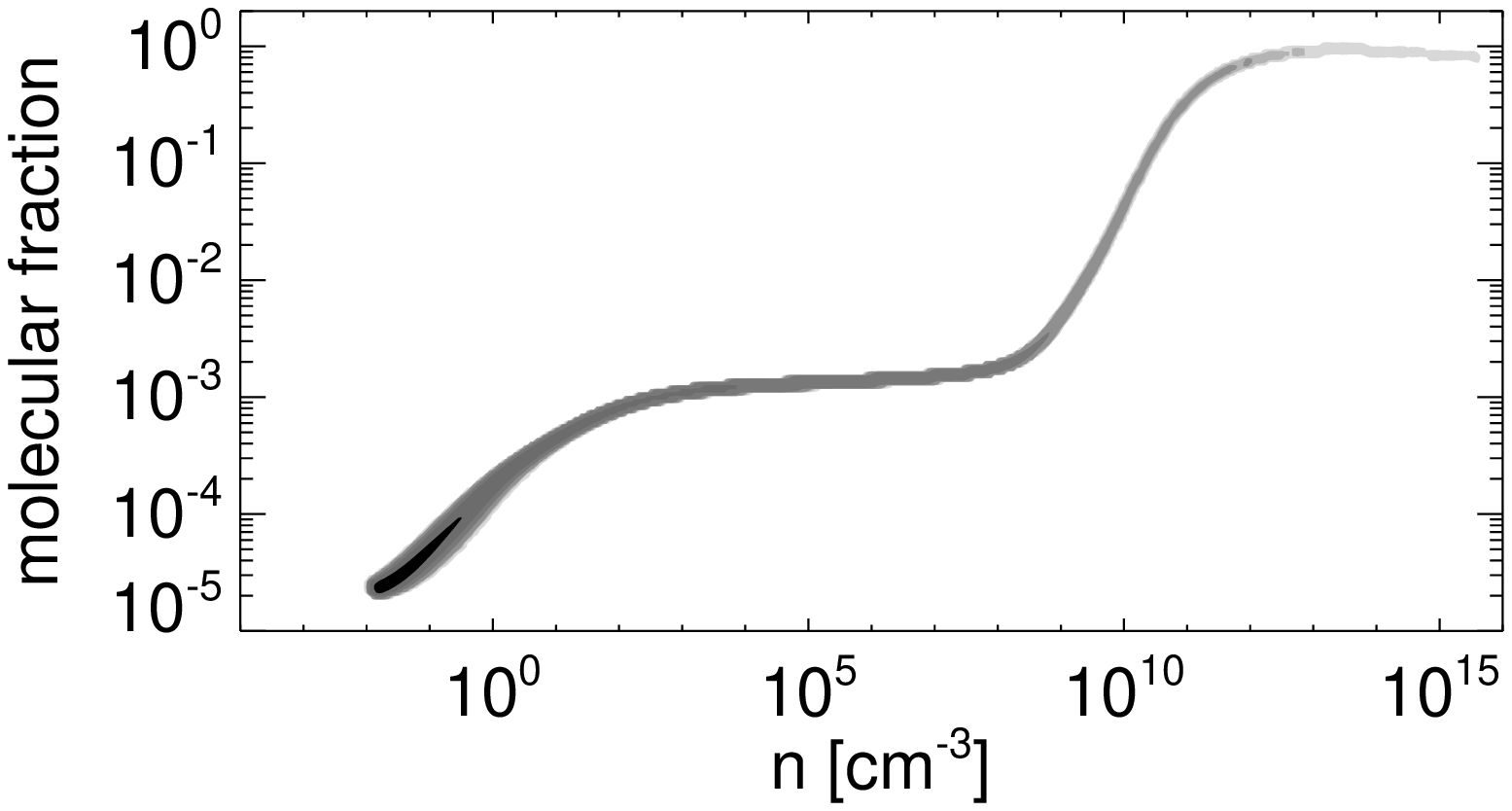}}
\caption{(Top) Gas distribution in the temperature-density phase space
in a spherical collapse problem.
The indicated characteristic features are explained as follows;
(A) gas temperature reaches $\ga 1000$K by virialization, 
and hydrogen molecules are formed by two-body processes,
(B) molecular hydrogen cooling brings the gas temperature down
to 200 K, (C) the \HH cooling rate saturates and becomes close
to the density-independent, LTE value, (D) three-body reactions kick in and the gas becomes
fully molecular, (E) the line cooling rate decreases as the density increases because of
the cloud's opacity, (F) collision-induced emission 
becomes a dominant cooling process, and (G) \HH dissociation begins at
$T \sim 2000$K. (Bottom) The molecular fraction $f_{\rm H2}$ of the gas.
The increase in the fraction at A, D, a plateau at C$\rightarrow$D, and 
the temporal decrease owing to dissociation at G are clearly seen in this plot. 
\label{fig:d_T}}
\end{inlinefigure}
 
Previous three-dimensional simulations of primordial gas cloud formation
were hampered by the complexity of calculating line opacities and the reduction
of the resulting cooling rate.
The maximum resolution, in terms of gas density, 
achieved in these simulations was thus $\sim 10^{10} {\rm cm}^{-3}$,
where the assumption of optically thin cooling breaks down.
With the novel technique described in Section \ref{sec:sobolev},
our simulations can follow the evolution of a primordial
gas cloud to $n_{\rm H} \sim 10^{16} {\rm cm}^{-3}$, nearly six orders
of magnitude greater than previous three-dimensional 
calculations reliably probed.
To study the detailed evolution of a proto-stellar ``seed'' 
beyond $n_{\rm H} \sim 10^{16} {\rm cm}^{-3}$ , 
we would need to implement a few more physical processes, as explained
in Section 3.5.

An important quantity we measure is the optically-thick line cooling 
rate, which serves as a critical check of our numerical implementation.
Fig. \ref{fig:freduce} shows the normalized \HH line cooling rate
against local density. We use an output at the time when the central 
density is $n_{\rm c}=10^{14} {\rm cm}^{-3}$. 
In the figure, we compare our simulation results with those from the full 
radiative transfer calculations of ON98 (open diamonds). 
Clearly our method works very well. 
The steepening of the slope at $n > 10^{12} {\rm cm}^{-3}$, owing to the
velocity change where infalling gas settles gradually onto
the center, is well-reproduced. We emphasize that the level of agreement 
shown in Fig. \ref{fig:freduce} 
can be achieved only if {\it all} of the local densities
(of chemical species), temperatures,
and velocities are reproduced correctly. 

\begin{inlinefigure}
\resizebox{10cm}{!}{\includegraphics{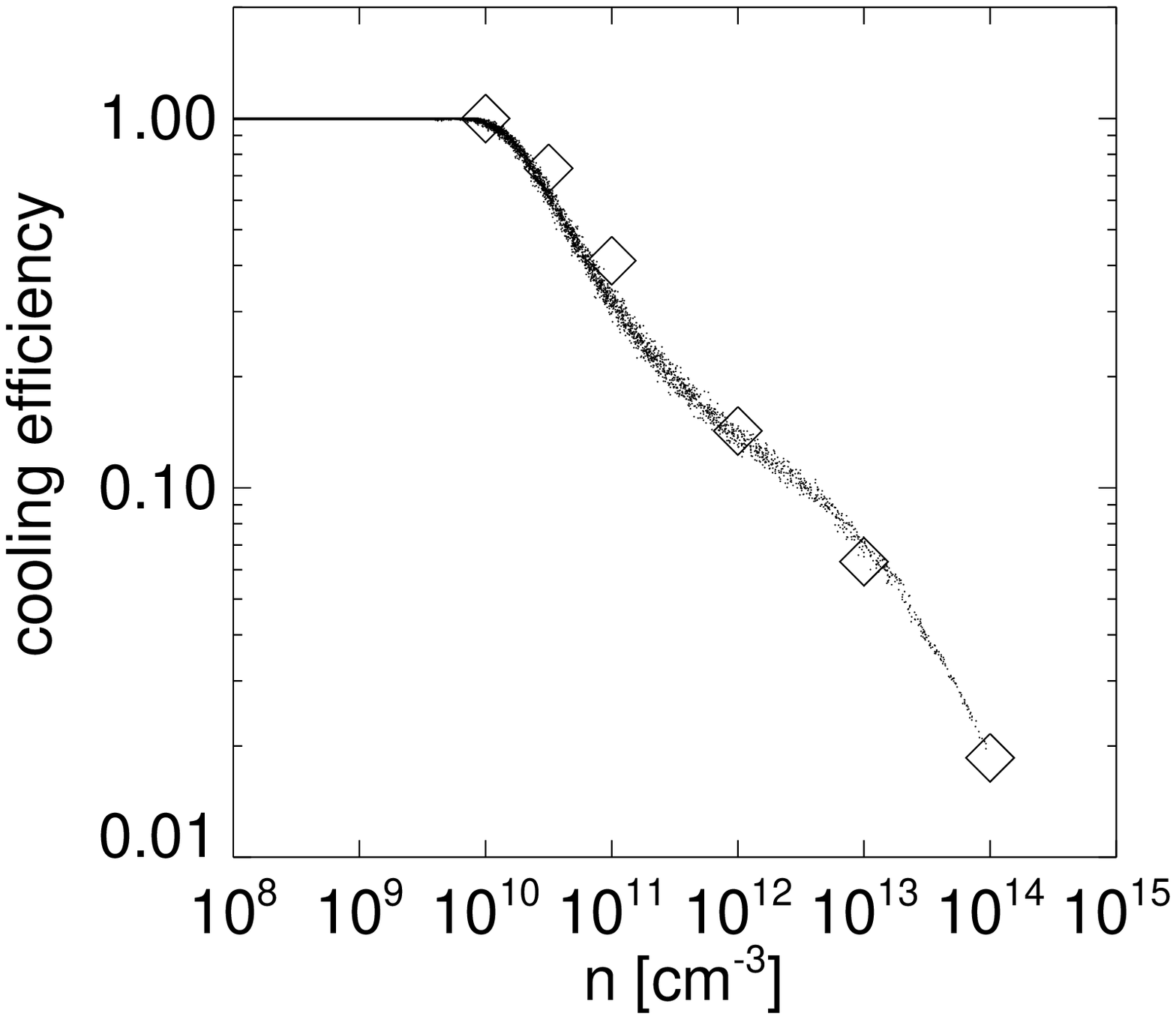}}
\caption{The cooling efficiency defined by 
$f=\Lambda_{\rm thick}/\Lambda_{\rm thin}$. The open squares 
are the results from the one-dimensional calculation of Omukai \& Nishi (1998).
We plot the efficiency as a function of local density at
the time when the central density is $n = 10^{14} {\rm cm}^{-3}$.
\label{fig:freduce}}
\end{inlinefigure}

\section{Cosmological simulations}

It is important to study the formation of primordial stars using a
proper set-up, starting from realistic initial conditions.  To this
end, we have run a
cosmological simulation adopting the concordance $\Lambda$CDM
cosmology with matter density $\Omega_{m}=0.26$, baryon density
$\Omega_{\rm b}=0.04$, cosmological constant $\Omega_{\Lambda}=0.7$,
and expansion rate at the present time $H_{0}=70$km s$^{-1}$ Mpc$^{-1}$.
The power spectra of the initial density fluctuations for baryons and
dark matter are calculated by the Boltzmann code of Sugiyama (1995).
In the examples discussed below, the power spectra were normalized to
$\sigma_8 = 0.9$, consistent with the first year WMAP results, but
slightly larger than the recent estimate of Spergel et al. (2006),
based on their third year data-set.  Everything else being equal, a 
smaller value of $\sigma_8$ will shift structure formation to slightly
lower redshifts.  The main purpose of our present investigation is to
study the physics of gas collapse to high densities and fragmentation 
in dark matter halos.  These processes will be unaffected by a shift
in $\sigma_8$, although the cosmic time corresponding to the
evolution we model would be different, as would the abundance of the
halos and stars at a given redshift.

In order to avoid spurious clumping in the initial particle set-up, we
use ``glass'' particle distributions (White 1996).  Further details on
the initial conditions are found in Yoshida, Sugiyama \& Hernquist
(2003).  We use a simulation volume of $0.3$ Mpc on a side.
Starting from a low resolution simulation, we select the most massive
halo at $z=15$ in the volume and apply a hierarchical zoom-in
procedure (e.g. Navarro \& White 1993; Tormen et al. 1997; Gao et
al. 2005) to the region surrounding the halo, so that progressively
higher mass resolutions are realized.  In the highest resolution part
of the the initial conditions, the dark matter particle mass is 0.0944
$M_{\odot}$ and that of gas particles is 0.0145 $M_{\odot}$.  In order
to achieve an even higher mass resolution, we progressively refine the
gas particles using the method of Kitsionas \& Whitworth (2002) and
Bromm \& Loeb (2003) as the gas cloud collapses. We do this refinement
so that a local Jeans length is always resolved by fives times the
local SPH smoothing length. With this technique, we achieve a mass
resolution of $m_{\rm p}=60 M_{\oplus}$, where $M_{\oplus}$ denotes
the Earth mass, at the last output time. For the study of cloud
fragmentation at intermediate densities $n_{\rm H} = 10^9-10^{11} {\rm
cm}^{-3}$, we carry out two simulations; one with the refinement
technique and the other without it, to verify that temporal
perturbations caused by refinement do not affect the results presented
in Section \ref{sec:stability}.

\vspace{1cm}
\begin{inlinefigure}
\resizebox{16cm}{!}{\includegraphics{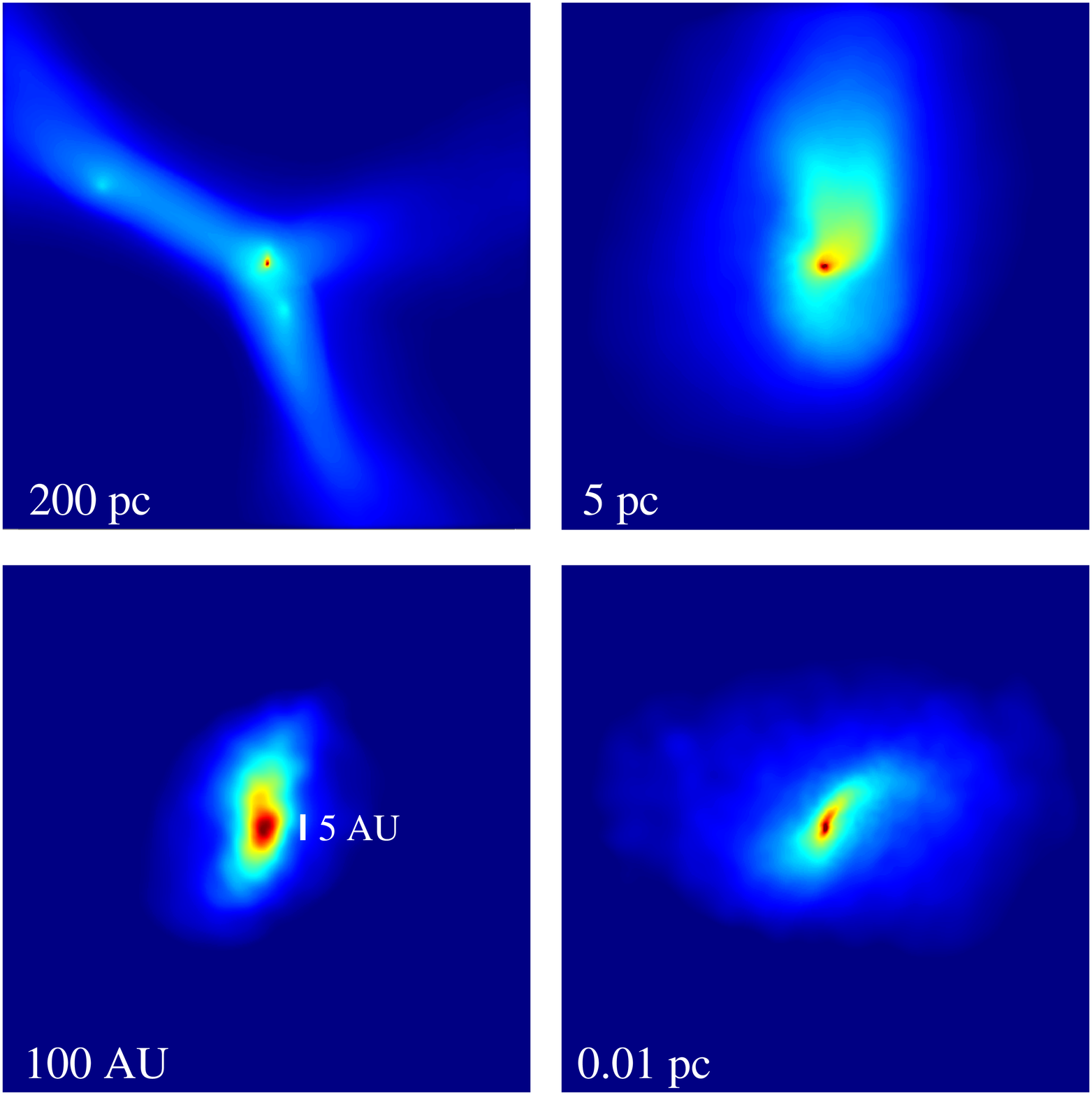}}
\caption{Projected density distribution for our cosmological simulations
at a redshift $z\approx 19$.
The physical side-length is indicated in each panel. 
\label{fig:cosmo}}
\end{inlinefigure}

\begin{inlinefigure}
\resizebox{16cm}{!}{\includegraphics{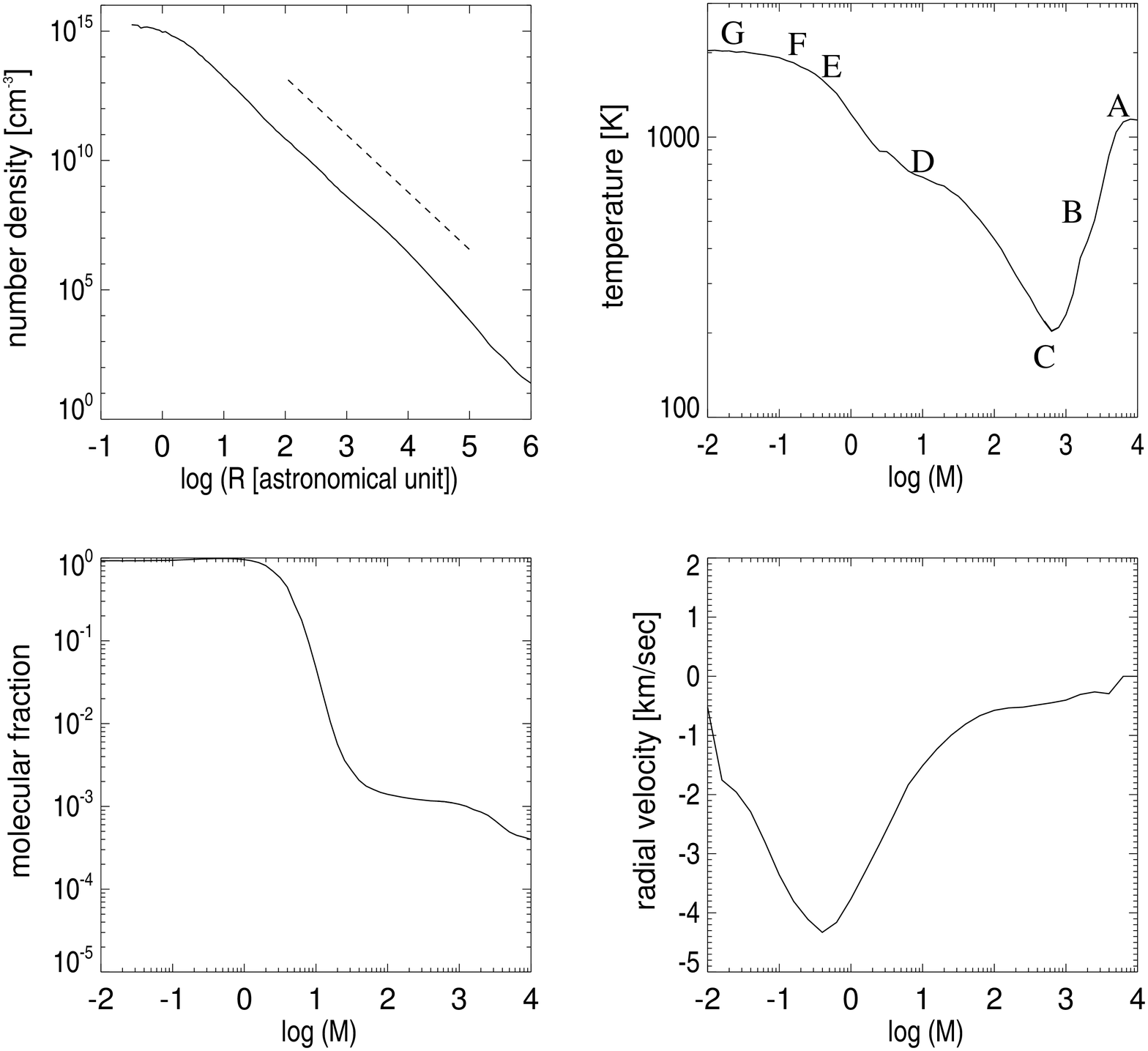}}
\caption{Radial profiles for density, temperature,
molecular fraction, and infall velocity
at a redshift $z\approx 19$.
The density is plotted as a function of distance from the center,
whereas the other three quantities are plotted as a
function of enclosed gas mass. The density profile is close to
the power-law  $\propto R^{-2.2}$. The characteristic features
in the temperature profile (marked A-G) can be explained as in
Fig. \ref{fig:d_T} (see also text). 
\label{fig:profiles}}
\end{inlinefigure}
 
\subsection{Thermal evolution of the proto-stellar gas cloud}
\label{sec:thermal}
In our $\Lambda$CDM simulation, the first gas cloud is formed at the center
of a dark matter halo with a mass of $6\times 10^5 M_{\odot}$.
The halo is located at the intersection of filamentary structures
at a high-density peak.
Fig. \ref{fig:cosmo} shows the projected gas density around the proto-stellar
gas cloud. At the output time, the density of the cloud core has reached 
$\sim 3\times 10^{15} {\rm cm}^{-3}$. 
The four panels in Fig. \ref{fig:cosmo} show progressively zoomed-in 
views of the central regions,
with the physical dimension indicated in each panel (clockwise from 200 parsec to 100 AU).
In the lower-left panel, we compare the size of the cloud core
with a solid bar which indicates a length of 5 astronomical units!
The central part in the top-right panel has a mass of $\sim 300 M_{\odot}$
with an approximate diameter of $\sim 1 {\rm pc}$; this region
is self-gravitating and undergoing
run-away collapse. In the bottom-right panel,
the central darkest area is nearly fully molecular with a mass
of about $1 M_{\odot}$. Finally the darkest region in the
bottom-left panel has a mass of $\sim 0.01 M_{\odot}$,
which is completely opaque to molecular lines 
and is cooling by CIE.

Fig. \ref{fig:profiles} shows the radial profiles of
density, temperature, velocity, and molecular fraction
at the final output time.
Except for the density profile which is shown as a function of radius
(note the x-axis is in astronomical units),
the mass coordinate (enclosed gas mass)
is used for the horizontal axis.
Although the plotted profiles are for a single output time,
the gas evolution can indeed be easily inferred from them,
because the collapse time scales as $\propto 1/\sqrt{n}$, 
leaving the profiles of the outer region 
almost unchanged while the central parts collapse further. 
The density profile evolves self-similarly
approximately as a power-law with $n \propto r^{-2.2}$
(dashed line in the top-left panel),
being consistent with previous one-dimensional
and three-dimensional simulations.

The characteristic features of the temperature profile are 
understood by various physical processes as we discussed already
in the spherical collapse simulation. 
Virialization brings the gas temperature to above 1000 K, 
and hydrogen molecules are formed by two-body reactions 
(point A in the top-right panel of Fig. \ref{fig:profiles}).
The cloud cools by \HH cooling (B) and the gas temperature
is lowered to about 200 K (C). There, the gas density is
$\sim 10^4 {\rm cm}^{-3}$,
beyond which the \HH rotational level population
approaches that of LTE. Then, the cooling rate per molecule becomes
density-independent, and radiative cooling is much less
efficient than in the low density limit.
This is the so-called loitering regime (Bromm et al. 2002),
and further collapse is induced dynamically
when the cloud becomes Jeans-unstable.
To see the onset of run-away collapse, we compare 
the enclosed gas mass with the locally estimated 
Bonnor-Ebert mass (Bonnor 1956; Ebert 1955):
\begin{eqnarray}
M_{\rm BE} &=& \frac{m_{1} c_{\rm s}^4}{G^{3/2} P_o^{1/2}},\nonumber \\
&\approx &20 M_{\odot} T^{3/2} n^{-1/2} \mu^{-2} \gamma^2 \, ,
\end{eqnarray}
where $m_1$ is the first maximum mass of the solution for the isothermal 
Lane-Emden equation (see, e.g. Stahler \& Palla 2004), and $\mu$
and $\gamma$
denote the mean molecular weight and adiabatic index, respectively.
We approximate the external pressure by its local value
taken from the radial density and temperature profiles.
(Note that $M_{\rm BE}$ evaluated in this manner is essentially
the same as the local Jeans mass.)
We find that the enclosed gas mass exceeds the
Bonnor-Ebert mass at $M \sim 200 M_{\odot}$. This is the
characteristic mass of the collapsing gas cloud.
We see in Fig. \ref{fig:profiles} that the temperature rises
below this mass scale, showing clearly the onset of collapse
and heating by contraction (Point C).
The $200 M_{\odot}$ cloud is now self-gravitating, being essentially
decoupled dynamically from the host dark matter halo.

Three-body reactions convert most of hydrogen into molecules (D in the top-right 
panel). As the molecular fraction profile (bottom-left panel) shows, the central 1-2$M_{\odot}$ 
becomes fully molecular through these rapid reactions. 
Cooling by \HH lines soon saturates because of
gas opacity (E). When the core density reaches $\sim 10^{14} {\rm cm}^{-3}$,
\HH line cooling becomes very inefficient, but CIE cooling kicks in (region F).
At higher densities, the gas temperature is maintained
at $\sim$ 2000K, whereby dissociation of hydrogen molecules effectively {\it absorbs}
heat input from contraction and dissipation of gravitational energy
(G). Dissociation already commenced in the innermost region at the plotted output time,
as is
clearly seen in the molecular fraction profile at $M_{\rm enc} <
0.1 M_{\odot}$.

\subsection{Thermal instability and fragmentation}
\label{sec:stability}
It is an outstanding question whether
or not primordial gas clouds are subject to thermal instability
and fragment into multiple objects.  Fragmentation could be
manifested in various forms because it depends on the cloud
geometry, thermal history, interaction with ambient matter, etc. 
Here, we focus on fragmentation of a primordial gas cloud
collapsing in a CDM halo, with properties as found in our cosmological simulation. 
In our simulation, we find that the central gas cloud does
not break up into multiple objects but remains intact
as a single entity until the last output time. 
To study cloud fragmentation in detail and more generally, we examine
the stability of the gas against thermal instability.

A well-known stability criterion is
the so-called Field criterion:
\begin{equation}
\rho_0 \left(\frac{\partial L}{\partial \rho}\right)_{T} 
- T_0 \left(\frac{\partial L}{\partial T}\right)_{\rho} < 0,
\end{equation}
where $L$ is the energy loss rate per unit mass (Field 1965).
Because the molecular hydrogen cooling rate
is density-independent at high densities,
and has a steep temperature dependence $\propto T^{\alpha}$ with
$\alpha > 1$ at $100 < T [{\rm K}] < 10000$ (see Fig. 1),
the Field criterion is satisfied for primordial gas
in this region {\it for a fixed molecular fraction}.
Thermal instability can also be driven by a change in  
chemical composition, such as ionization/recombination
(Defouew 1970; Goldsmith 1970) or molecule formation/dissociation
(Yoneyama 1973).  
Sabano \& Yoshii (1977) and Silk (1983) suggest
that primordial gas can be chemo-thermally unstable.
Chemo-thermal instability can be triggered when a rapid increase
in the coolant fraction (\HH molecules) induces
efficient cooling and condensation, leading to an enhanced
production of molecules via three-body reactions.  
There are two proposed regimes where the chemo-thermal instability
can be triggered: one at $n \sim 10^{10} {\rm cm}^{-3}$ where
three-body reactions make the overall gas cooling very efficient,
and the other at  $n \sim 10^{15} {\rm cm}^{-3}$
where collision-induced emission cooling comes into play.

Previously, ABN02 concluded that gas clouds
do not fragment because of efficient turbulent mixing. 
Using one-zone calculations, Omukai \& Yoshii (2003) 
conclude that the cloud core becomes unstable,
but does not fragment because the cloud is already
compact. Ripamonti \& Abel (2004) study the thermal 
instability in the high density regime and argue
that fragmentation is unlikely to occur. 
None of these previous works provide definitive answers,
unfortunately. The high-resolution simulation of ABN02 
employs the optically-thin approximation for \HH cooling,
and thus their results do not provide accurate estimates
for relevant quantities at densities above $n>10^{10} {\rm cm}^{-3}$, 
whereas the other studies based on zero- or one-dimensional
calculations suffer from the fact 
that fragmentation is intrinsically a three-dimensional problem.
We are able to, for the first time, directly address this issue
using a three-dimensional simulation with sufficient mass-
and ``physical'' resolution in a fully cosmological
context. 

We follow Omukai \& Yoshii (2003) to investigate the stability
of the gas cloud core against isobaric perturbations.
For perturbations in temperature, density, and
molecular fraction, $\delta T, \delta \rho, \delta x
\propto \exp(\omega t)$, under the constraint 
$\delta P = \delta T + \delta \rho + \delta x = 0$,
linear stability analysis yields the dispersion relation 
\begin{equation}
A \omega^2 + B \omega + C = 0,
\label{eq:dispersion}
\end{equation}
where
\begin{equation}
A = 1+\frac{6-f}{2(2-f)},
\label{eq:coeffA}
\end{equation}
\begin{equation}
B = \frac{\mu m_p}{kT}(TL_{T}-\rho L_{\rho}-L)
-\frac{\mu\chi}{2kT}(TF_{T}-\rho F_{\rho} - F)
- A\left(F_{f}+\frac{\mu}{2}\rho F_{\rho} +\frac{\mu F}{2}\right),
\label{eq:coeffB}
\end{equation}

\begin{inlinefigure}
\resizebox{10cm}{!}{\includegraphics{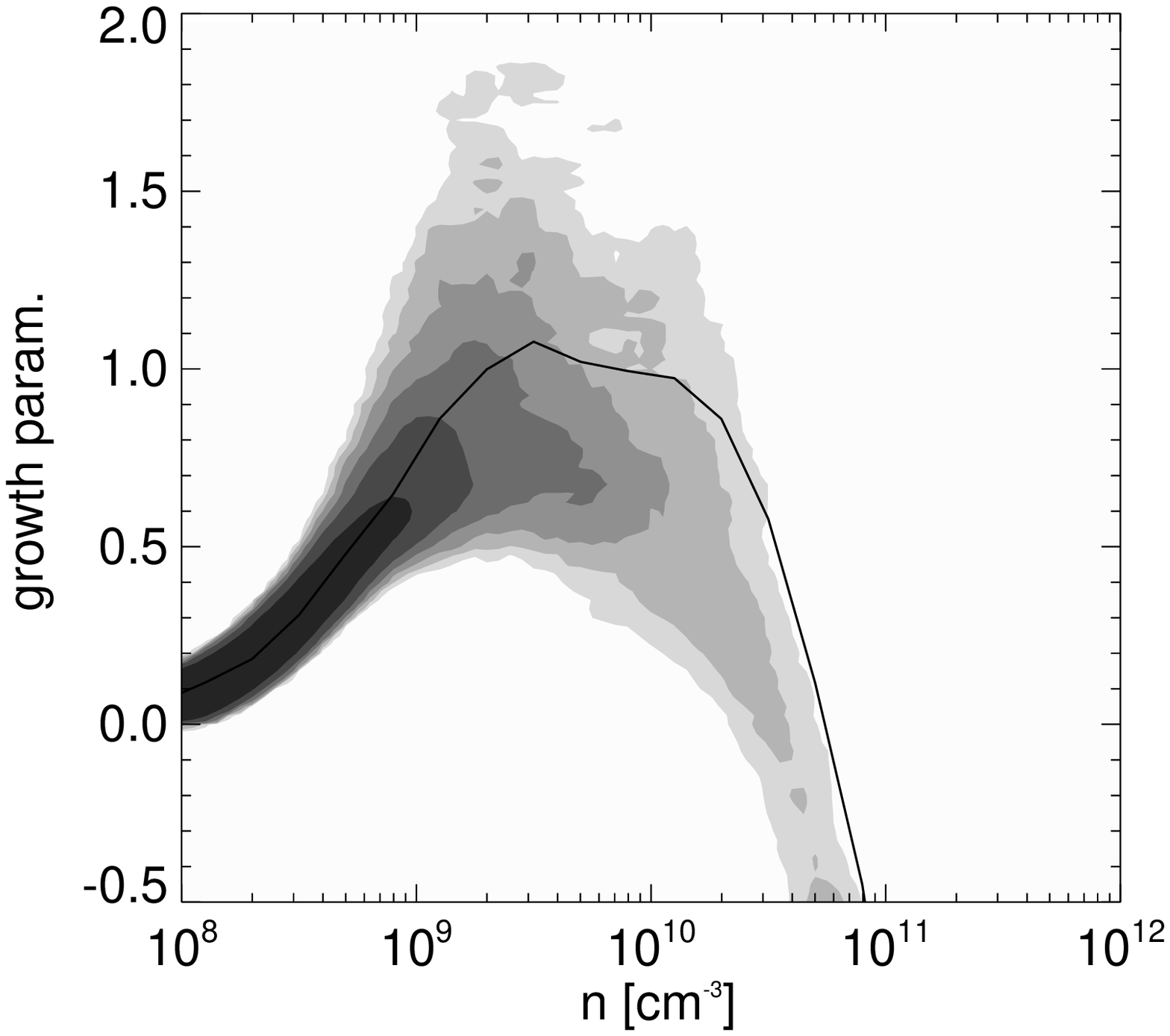}}
\caption{The instability growth parameter $Q$ as
a function of gas density. The shaded contour shows
the distribution of the cloud core gas at a
time when the central density is $\sim 10^{12} {\rm cm}^{-3}$.
The thick solid line shows the evolution of the
most dense particle in this $Q-n$ plane. 
\label{fig:frag_low}}
\end{inlinefigure}

\noindent and
\begin{eqnarray}
C &=& -\frac{\mu m_p}{kT}(TL_{T}-\rho L_{\rho}-L)
\left(F_{f}+\frac{\mu}{2}\rho F_{\rho} + \frac{\mu F}{2}\right)\nonumber \\
&+&\frac{\mu m_p}{kT}
\left(L_f + \frac{\mu}{2}\rho L_{\rho}+\frac{1}{6-f}L\right)
(TF_T -\rho F_{\rho})\nonumber \\
&+&\frac{\mu}{2}F
\left[\frac{\mu}{3-f/2}\left(\frac{1}{2}+\frac{\chi}{kT}\right)(TF_T-\rho F_{\rho}) 
-\frac{\chi}{kT} 
\left(F_f+\frac{\mu}{2}\rho F_{\rho}+\frac{\mu F}{2}\right)\right] \nonumber \\
&-&\frac{\mu}{(3-f/2) \;t_{\rm dyn}}(TF_T-\rho F_{\rho}).
\label{eq:coeffC}
\end{eqnarray}

\noindent In the above expressions, $f$ is the molecular fraction ($f=1$ for a fully molecular gas),
$\mu = 2/(2-f)$, $F = df/dt$ is the net formation rate of hydrogen molecules,
and $L$ is the rate of energy loss per unit mass. Subscripts denote the respective 
partial derivatives. We have defined the local dynamical time
\begin{equation}
t_{\rm dyn} = \sqrt{\frac{3\pi}{32 G\rho}}.
\end{equation}

The dispersion equation (\ref{eq:dispersion}) 
has a positive real root (a growing mode solution) if and only if $C < 0$,
because $A$ is positive by definition. 
Furthermore, for perturbations to grow in a gravitationally collapsing gas, 
the characteristic growth timescale $t_{\rm g} \propto 1/\omega$ must be 
shorter than the dynamical time $t_{\rm dyn}$. 
Hence, we define the growth parameter
\begin{equation}
Q=\frac{t_{\rm dyn}}{t_{\rm g}}=\omega t_{\rm dyn}.
\end{equation} 
If $Q > 1$, perturbations are expected to grow faster than
the gravitational contraction of the cloud (Omukai \& Yoshii 2003;
Ripamonti \& Abel 2004). 

Fig. \ref{fig:frag_low} shows the growth parameter
evaluated locally for all the gas elements at the time when the central density is
$n_{\rm c} = 10^{12} {\rm cm}^{-3}$.
We also show the evolutionary track for the cloud core
by the solid line.
We see that a part of the gas cloud is in the unstable
region where  $Q > 1$.
However, $Q$ never becomes much larger than unity. 
It is indeed less than 1.5 for almost all the gas.
Omukai \& Yoshii (2003) argue that $Q$ must be significantly
larger than unity for a gas parcel to break up into multiple objects.
This is because the size of the cloud, 
$l_{\rm c} \sim c_{\rm s} t_{\rm dyn}$, must be much larger
than the length of growing perturbations
$l_{\rm p} \sim c_{\rm s} t_{\rm g}$. 
In other words, while a gas parcel can
become thermally unstable, for the gas to break up
into multiple objects, $Q$ must be at least larger
than two so that the perturbations can grow before the region
gravitationally contracts. Fig. \ref{fig:frag_low} shows
that the cloud center never enters the regime $Q>2$.
Our results therefore
support the conjecture of Omukai \& Yoshii
that {\it the cloud becomes chemo-thermally unstable
but evolves into a single object, because the cloud 
is already too compact.}

 We do the same analysis at the time when the cloud core enters 
the high-density regime.
Evaluating equations (\ref{eq:dispersion})-(\ref{eq:coeffC}),
we find that the gas is always stable. To see this
more clearly, we show the gas distribution in a
thermodynamic phase diagram in Fig. \ref{fig:frag_high}.
We indicate the region where $Q>0$ by grey contours.
For estimating the value of $Q$ as a function of density
and temperature, we calculate $f_{\rm H_2}$ assuming
chemical equilibrium. We solve the cubic equation
for the equilibrium molecular fraction
\begin{equation}
k_{22} (1-f)^3 n^3+\frac{1}{2} k_{23} f(1-f)^2 n^3
= \frac{1}{2} k_{11} f(1-f) n^2 + \frac{1}{4} k_{24} f^2 n^2,
\end{equation}
where $k_i$ denote respective reaction rates tabulated
in the Appendix.
While the assumption of chemical equilibrium
does not strictly hold in all the plotted range of
density and temperature, it is indeed
a good approximation for the relevant region
where the actual gas evolutionary track lies.
As clearly seen in Fig. \ref{fig:frag_high},
the gas never approaches an unstable region.
It appears that, since the cooling rate is larger at higher density
and/or at higher temperatures (see the temperature dependence of
CIE cooling in Fig. \ref{fig:CIE_rate}), the gas evolutionary track 
{\it circumvents} the instability region.
It is implausible that a gas parcel enters the instability
region from the lower temperature side,
because the cause of the instability is radiative cooling,
which always tend to bring the gas temperature downward.
We further argue that, because dissociation of hydrogen molecules
occurring at $T>2000 $K effectively absorbs the
heat input by contraction and thermalization,
the gas evolves approximately isothermally
around $T\sim 2000 $K until full scale
dissociation is completed, and the temperature cannot increase 
much without the density increasing.
We therefore conclude that fragmentation owing to thermal
instability does not take place in this regime either.

\begin{inlinefigure}
\resizebox{10cm}{!}{\includegraphics{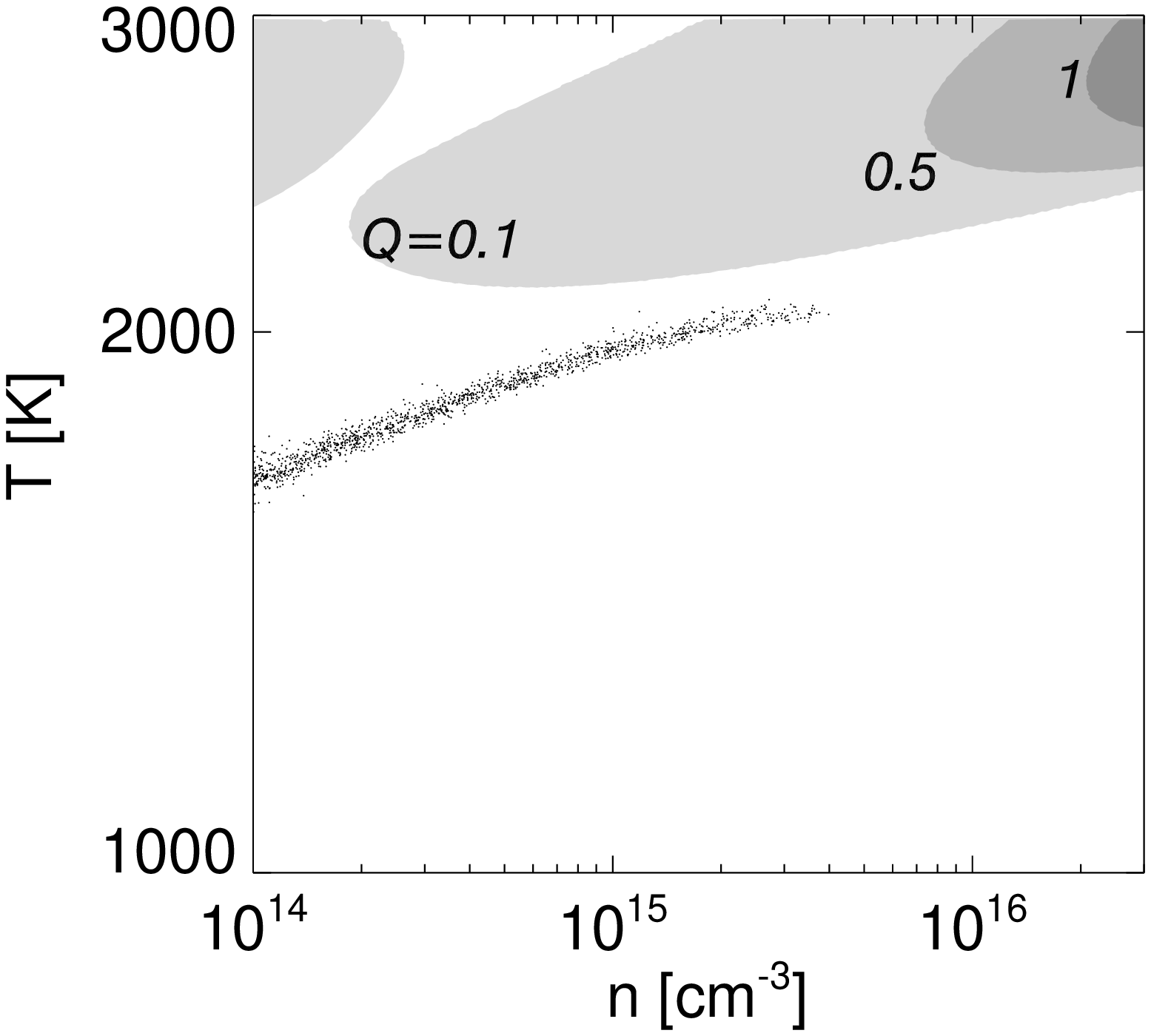}}
\caption{Stability analysis in the high-density regime.
We plot the distribution of gas in the cloud core 
in the temperature-density phase space. The shaded regions 
indicate regions with $Q=0.1, 0.5, 1$ from light 
to dark grey, respectively. The gas does not enter the
instability region.
\label{fig:frag_high}}
\end{inlinefigure}

\subsection{Stability against gravitational deformation}
Gas fragmentation could be triggered if a cloud is significantly 
flattened (Tohline 1981; Miyama et al. 1984; Tsuribe \& Inutsuka 1999;
Bromm, Coppi, Larson 1999) 
or if it contracts into filamentary structures 
(Uehara et al. 1996; Nakamura \& Umemura 2001).
Although Bromm et al. (1999) present a particular case which yields
a disk-structure, their simulation is set up {\it as such}, 
by imposing an initial spin. 
In cosmological simulations, primordial gas clouds
with such shapes are not generally found (Abel et al. 2002; 
Yoshida et al. 2003). 
There are two important mechanisms that lead to
cloud fragmentation; one is the growth of deformation
during collapse and the other is rotation-induced 
fragmentation. While these two mechanisms are, in general,
coupled to each other, we here examine each effect separately.
We first show that the dense, collapsing gas cloud formed with 
$n_{\rm H} \sim 10^{4} {\rm cm^{-3}}$ would not fragment again 
at higher densities.
The collapse of the dense core proceeds in a self-similar manner.
This so-called Larson-Penston-type self-similar solution 
is known to be unstable to non-spherical deformation if 
$\gamma= d{\rm log}P/d{\rm log}\rho$ 
is less than the critical value, $\gamma_{\rm crit}=1.097$ 
(Hanawa \& Matsumoto 2000; Lai 2000).
An unstable core elongates by this instability and eventually fragments.
Let us define the core's elongation as ${\cal E} \equiv (b-a)/a$, where $a$ 
and $b$ are short and long axis lengths, respectively.
The growth rate of the elongation $\nu = d{\rm log}{\cal E}/d{\rm log}{\rho}$
is calculated by Hanawa \& Matsumoto (2000) and Lai (2000) using linear 
perturbation theory.

\begin{inlinefigure}
\resizebox{10cm}{!}{\includegraphics{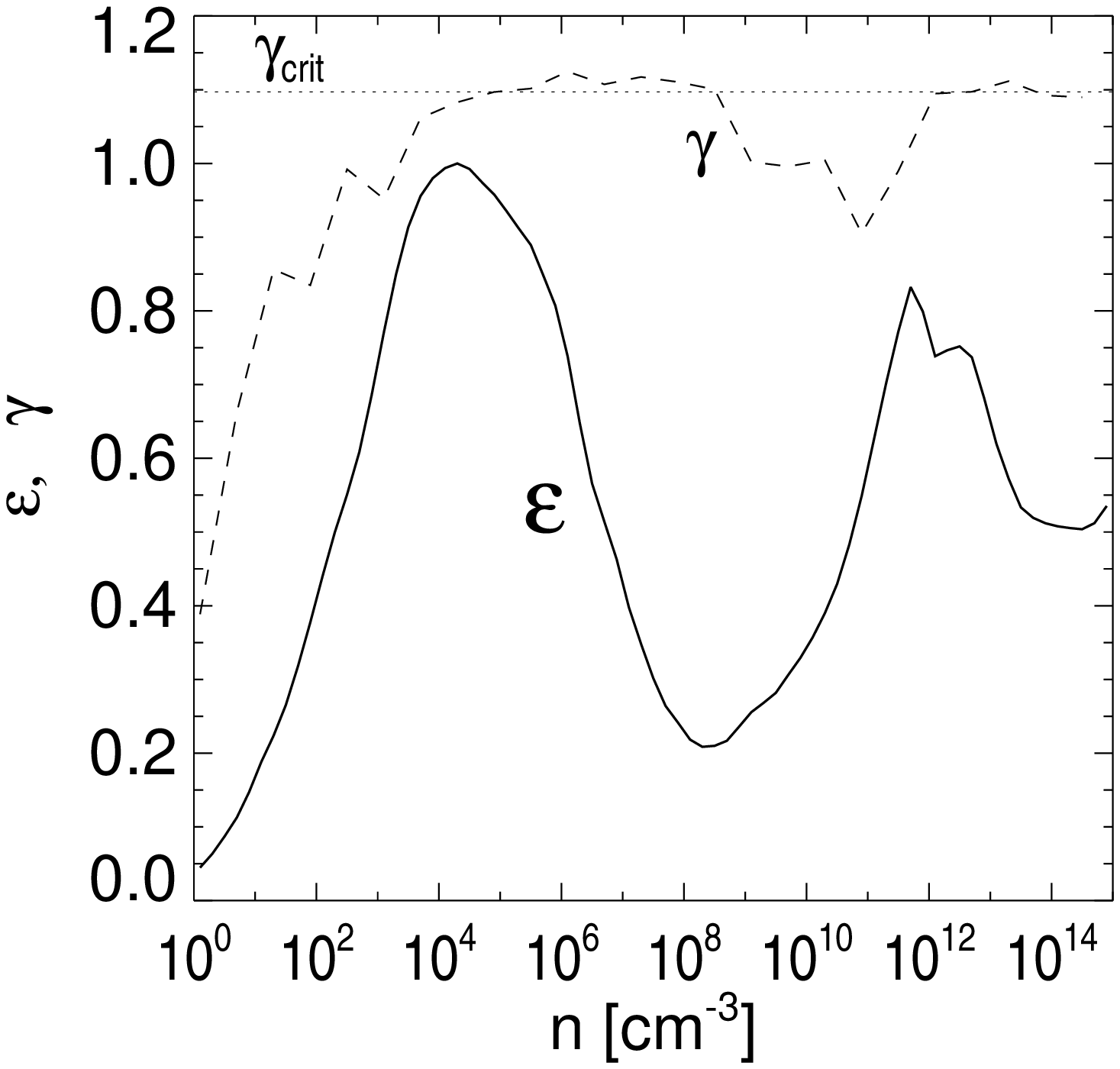}}
\caption{Linear stability analysis for deformation growth.
We plot the {\it predicted} evolution of the elongation ratio 
${\cal E}$ for the core
(solid line). The evolutionary track of the core stays below ${\cal E} = 1$
after the first run-away collapse at $n_{\rm H}\sim 10^4 {\rm cm}^{-3}$,
suggesting that the core is stable against deformation
to filamentary structure. 
The dashed line shows the evolution of the effective
ratio of specific heats, and the dotted line is the critical 
value $\gamma_{\rm crit}=1.097$ from linear theory.
\label{fig:elong}}
\end{inlinefigure}

\noindent Using the simulation outputs, we calculate the effective ratio 
of specific heats 
$\gamma= d{\rm log}P_{\rm c}/d{\rm log}\rho_{\rm c}$ 
at the density maximum as a function of density.
The evolution of the core elongation is then calculated as   
\begin{equation}
{\cal E}={\cal E}_{0} \;{\rm exp} \left( \int_{n_{\rm H,0}}^{n_{\rm H}} 
\nu(n_{\rm H}')\;{\rm d}\ln n_{\rm H}' \right)
\end{equation}
for the initial elongation ${\cal E}_{0}$ at $n_{\rm H,0}$.
In Fig. \ref{fig:elong}, we show the evolution as a function 
of the central number density.
The initial state is taken at the maximum of ${\cal E}$ around 
$10^{4} {\rm cm^{-3}}$, where the first run-away collapse begins.
Early in the evolution ($n_{\rm H} < 10^{9} {\rm cm^{-3}}$), 
$\gamma$ exceeds $\gamma_{\rm crit}$, and the elongation decays; 
the core becomes rounder.
With the onset of three-body H$_2$ formation, $\gamma$ starts
decreasing and falls slightly below $\gamma_{\rm crit}$.
There, the elongation grows but only slowly. As is seen in Fig. \ref{fig:elong},
it at most recovers the initial value ${\cal E}_{0}$. 
Since the effective $\gamma$ gradually increases
again at $n > 10^{10} {\rm cm}^{-3}$ when \HH line cooling becomes inefficient,
the deformation growth rate decreases.
We thus conclude that the core must be stable initially and remains stable to 
this mode of fragmentation throughout the collapse,
as is indeed the case in our simulation (see Fig. \ref{fig:cosmo}).

Next, we examine the effects of rotation. 
Tohline (1981) and many subsequent works (e.g. Tsuribe \& Inutsuka 1999)
conclude that basically two parameters, the cloud's initial ratio of thermal 
to gravitational energy ($\alpha_0$) and the initial ratio of rotational to gravitational
energy ($\beta_0$), determine whether or not the cloud will eventually fragment, 
for a given adiabatic exponent $\gamma$. 
Here, the parameters are defined as, respectively,
\begin{equation}
\alpha_0 = \frac{5 c_{\rm s}^2 R_0}{2GM}, \;\;\; 
\beta_0 = \frac{\Omega_0^2 R_0^3}{3GM}.
\end{equation}
Note that the latter quantity is essentially the degree of rotation support
(see Section \ref{sec:am} below).
Tsuribe (2002) investigated a particular case for $\gamma = 1.1$,
which is close to the actual value for primordial gas,
and concluded that the stability criterion is roughly given by
$\alpha_0 > 0.3$ for $0< \beta_0 < 0.3$. (Note that the stability region
does not differ much for a gas cloud with a constant
initial density or for a gas cloud with central condensation [Tsuribe 2002].) 
We calculate these quantities at the time
when the first run-away collapse begins,
namely when the core density is $\sim 10^4 {\rm cm}^{-3}$
and the temperature is $\sim 200$K (Point C in Fig. \ref{fig:profiles}).
We find $\alpha_c = 1.0$ and $\beta_c = 0.1$, which satisfies the stability 
condition. In summary, the gas cloud in our cosmological simulation
is found to be stable against deformation and rotation-induced fragmentation.
We have validated the numerical results using analytical stability criteria. 

\clearpage
\begin{inlinefigure}
\resizebox{10cm}{!}{\includegraphics{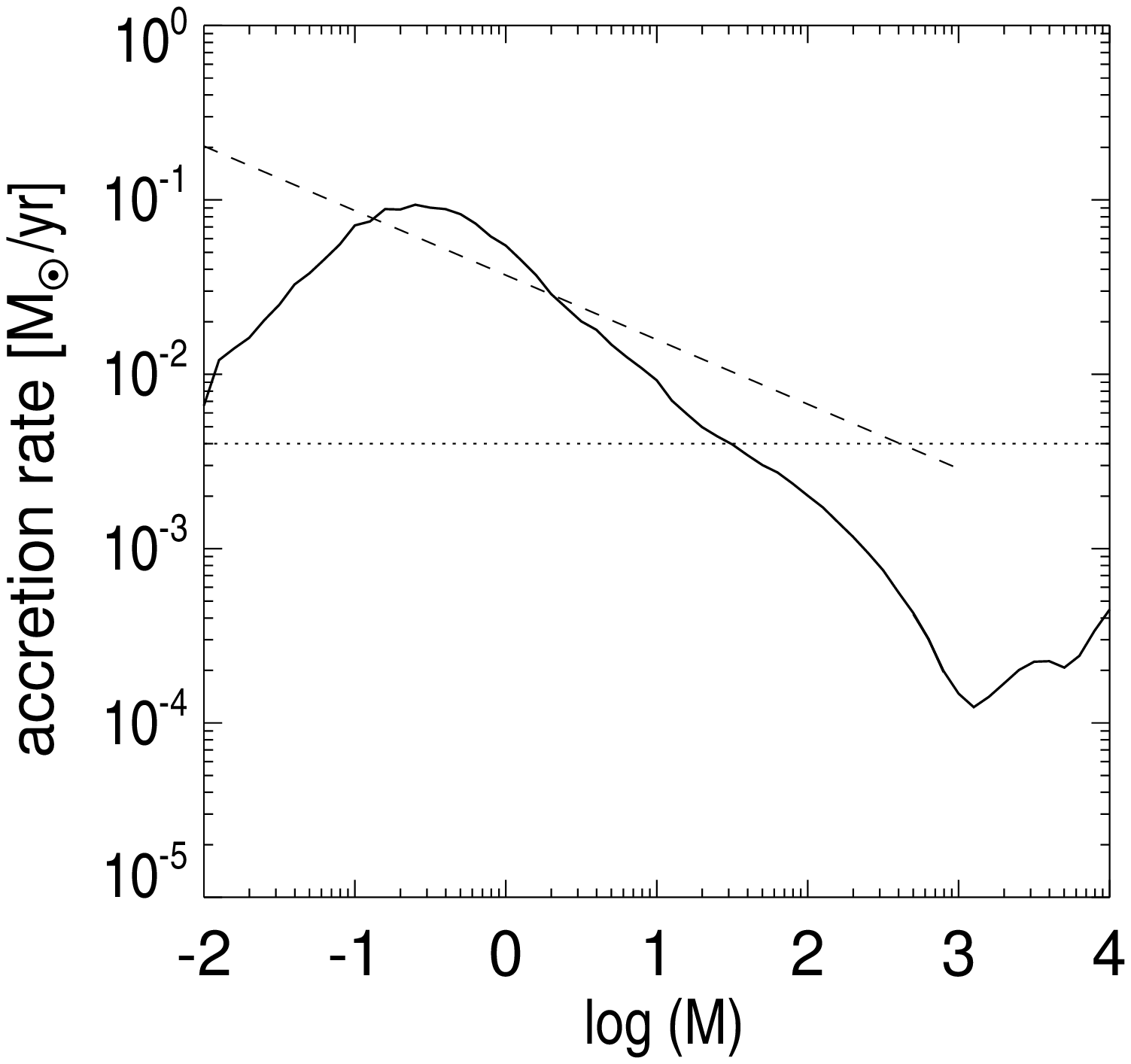}}
\caption{Instantaneous gas mass accretion rate
around the proto-stellar core (solid curve).
We evaluate equation (\ref{eq:massacc}) and plot it
as a function of enclosed gas mass. For reference,
the dashed line shows the accretion rate obtained by Omukai \& Nishi 
(1998) using their post-singularity, self-similar solution.
The dotted line is the critical mass accretion rate given by 
equation (\ref{eq:critacc}).
\label{fig:massacc}}
\end{inlinefigure}

\subsection{Gas mass accretion rate}
We have seen that the prestellar cloud does not fragment into multiple
clumps. Since the entire cloud has a much larger mass than 
the proto-stellar ``seed'', the protostar will evolve 
in an inside-out manner, by accreting a large quantity of surrounding gas. 
The rate of accretion of the infalling gas
is among the most important quantities in proto-stellar
evolution. Since the late time evolution of the protostar
itself affects the matter accretion rate,
we are not able to measure the 
true mass accretion rate onto the protostar directly.
Nevertheless, the instantaneous accretion rate
likely provides a good estimate.
In Fig. \ref{fig:massacc}, we show the instantaneous
gas mass accretion rate at the last output time
\begin{equation}
\dot{M} = 4\pi \rho\;r^2 v_{\rm rad} (r)
\label{eq:massacc}
\end{equation}
using again the mass coordinate.
We compare our result with that of ON98 
which is derived from the post-singularity
self-similar solution (Yahil 1983).
The model accretion rate
of ON98 has a shallower slope (i.e., larger accretion rate
in the outer regions) than that of our simulation.
This is reasonable because the ON98 model is
based on the Larson-Penston type self-similar solution
that predicts higher infall velocities at large distances
because of the assumption of the solution itself;
the velocity difference owes to the initial conditions
and boundary
effects, which are self-consistently calculated in our simulation.

A simple, order of magnitude estimate
for the mass accretion rate is given
as a function of sound speed as
\begin{equation}
\dot{M} \sim \frac{c_{s}^3}{G},
\label{eq:Mdot_estimate}
\end{equation}
and thus it is large for a high temperature gas.
For $T \sim 1000$ K,
the above estimate yields $\sim 10^{-3} M_{\odot} {\rm yr}^{-1}$,
similar to the simulation result. 
It is illustrative to compare this value with that in 
present-day star-forming regions which have $T \sim 10$ K.
The temperature difference naturally explains
the much larger accretion rate, by more than two orders of magnitude, 
for the primordial case.
Intriguingly, the overall feature of the accretion rate
shown in Fig. \ref{fig:massacc} can be
explained qualitatively from the temperature profile.
The gradual increase at $M_{\rm enc} = 100-1000 M_{\odot}$
corresponds to the gradual temperature increase shown in the top-right panel in Fig. \ref{fig:profiles}.
The slight shallowing at $M_{\rm enc} \sim 10 M_{\odot}$
corresponds to the temperature ``dip''
where three-body reactions promote molecule formation.
Inside the radius, $M_{\rm enc} < 10 M_{\odot}$, 
the temperature rises again, and the accretion rate increases
as well. The decline at the inner-most part is simply
because of the output timing. At this time, the radial velocity 
gradually decreases towards the center (Fig. \ref{fig:profiles});
i.e. a radiative shock has not yet formed.
In the outer part, $M_{\rm enc} > 1000 M_{\odot}$,
the increasing accretion rate owes to gravitational pull 
by the host dark matter halo,
which is not seen in cloud evolution calculations without dark 
matter.

The instantaneous accretion rate is larger 
than ABN02 found over the entire plotted mass range.
While the small accretion rate at $M_{\rm enc} < 10 M_{\odot}$
in ABN02 can be understood as a result of over-cooling
(see Fig.2 in ABN02), it is unclear why the large-scale 
($M_{\rm enc} \ga  100 M_{\odot}$) mass accretion rates
differ. It cannot be attributed to the degree of
rotational support because we obtained quite similar 
rotation values to ABN02 (see the next section).
To study this discrepancy further, we carry out two 
controlled simulations by embedding a primordial gas within 
an NFW dark matter halo similarly to the simulation
presented in Section \ref{sec:sphere}. 
For one case we set $f_{\rm b} = 0.16$, consistent
with our $\Lambda$CDM cosmology, whereas
for the other case, we set $f_{\rm b} = 0.05$,
that is the value adopted in the standard cold
dark matter simulation with $\Omega_{\rm m}=1$
of ABN02. For both cases we assign zero velocities initially 
and let the gas cloud collapse in the halo's potential 
with the same mass. We measured the resulting mass accretion 
rates at the time when the central density reaches
$n_{\rm c}=10^{15} {\rm cm}^{-3}$.
We find, interestingly, that the low baryon fraction case agrees 
well with the result of ABN02. The overall cooling time 
in the low baryon fraction case is longer
than in the other one and hence collapse takes place more slowly. 
In the high baryon fraction case, external pressure from the 
ambient gas at the onset of collapse may also affect
the accretion rate (Hennebelle et al. 2004; Motoyama \& Yoshida 2003).
Given the insignificant difference of up to a factor of a few,
we do not discuss the discrepancy further in the present paper.
Recently, O'Shea \& Norman (2006a,b) performed a series of cosmological
simulations to examine various environmental effects, such
as large-scale local density and halo formation epoch, on the 
instantaneous gas mass accretion rate. They indeed find a substantial 
variation in accretion rate among different halos. 
We mention that very different cosmological environments,
such as those realized in the early large-scale structure simulation
of Gao et al. (2005) could have a noticeable impact
on the mass accretion rate.

\subsection{Angular momentum}
\label{sec:am}
The gas cloud starts collapsing with a finite amount of 
angular momentum (AM), and thus
it is expected to spin-up gradually as it contracts,
unless there are some mechanisms to transfer AM.
Rotational support of the gas may play an important role in the process
of accretion. To first order, we expect that mass accretion
continues only if rotation does not halt collapse; 
i.e. if the angular momentum of the gas is small.
We calculate the specific angular momentum profile
around the protostar at the final output time. 
Fig. \ref{fig:AM} shows the profile
as a function of enclosed gas mass. The power-law profile is 
typical for gravitationally collapsed objects (Bullock et al. 2001;
van dens Bosch et al. 2002). The AM profile's shape and even the amplitude are
quite similar to the one found in the simulation of ABN02 over the entire
plotted range.
The degree of rotational support defined as
\begin{equation}
f_{\rm r} = \frac{L_{\rm sp}/r}{V_{\rm kep}},\;\;\;V_{\rm kep}=\sqrt{\frac{G M}{r}},
\label{eq:rot}
\end{equation} 
is shown by the dashed line. Here, $L_{\rm sp}$ is the specific AM
of the gas within radius $r$, and $V_{\rm kep}$ is the Kepler velocity
at that radius. It is about 40-50\% for
$M_{\rm enc}=0.01 - 1000 M_{\odot}$, and rotation is not (yet) able to stop 
the gas inflow. We have also measured $f_{\rm r}$ at previous 
output times and found that $f_{\rm r}$ has been monotonically increasing
at small radii.
To understand the reason for the characteristic AM profile,
i.e., why low angular momentum material resides 
in the central regions, we examine two quantities.

First, we make a conjecture that the region with
little angular momentum collapses first.
By measuring the probability distribution of specific AM
in the cloud before the onset of collapse, we show that 
this is indeed the case in our simulation.
At a time when the maximum density reaches $10^4 {\rm cm}^{-3}$,
we define the ``core'' as the region where the density
is larger than half that of the center. The center is defined to be the position 
of the most dense particle.
We then calculate the specific angular momentum
with respect to random points within the core,
by taking a fluid parcel of mass $0.1 M_{\rm Jeans, core} = 30 M_{\odot}$
around the points. The resulting probability
distribution is shown in Fig. \ref{fig:AMevo}.
There, the arrow indicates the value for the ``true'' center
that collapses first and becomes the densest part
{\it in the later output time}.
We interpret this, together with the specific AM distribution,
as evidence that regions with 
low angular momentum collapse fast in the (weakly-)turbulent
medium. There is alway a central core 
within which density fluctuations are erased by 
soundwaves (hence having a size $\sim c_{\rm s} t$), 
and there is also an AM distribution within it.

We further examine the evolution of AM in detail,
making use of the Lagrangian nature of the SPH
method. 
We trace the inner-most particles and calculate the 
evolution of the specific AM. We mark gas particles 
according to their mass coordinate at the final output time
and divide them into four shells.
We then keep track of the positions and velocities 
of the marked particles at earlier output times.
At each output time, we shift the spatial coordinate
such that the ``true'' central particle is always at 
the origin of the coordinate. (We note that the specific AM
profile varies considerably if we define the center as the
most dense region at each output time.)
Fig. \ref{fig:AMevo} shows the evolution of the specific angular 
momentum for the four mass shells.
As is clearly seen in the right panel of Fig. \ref{fig:AMevo},
the specific angular momentum of each mass shell is
well-conserved, while the mean radii decreased by
a factor of 3-5. This causes the gradual increase of
rotational support in the inner region as discussed previously.

\begin{inlinefigure}
\resizebox{10cm}{!}{\includegraphics{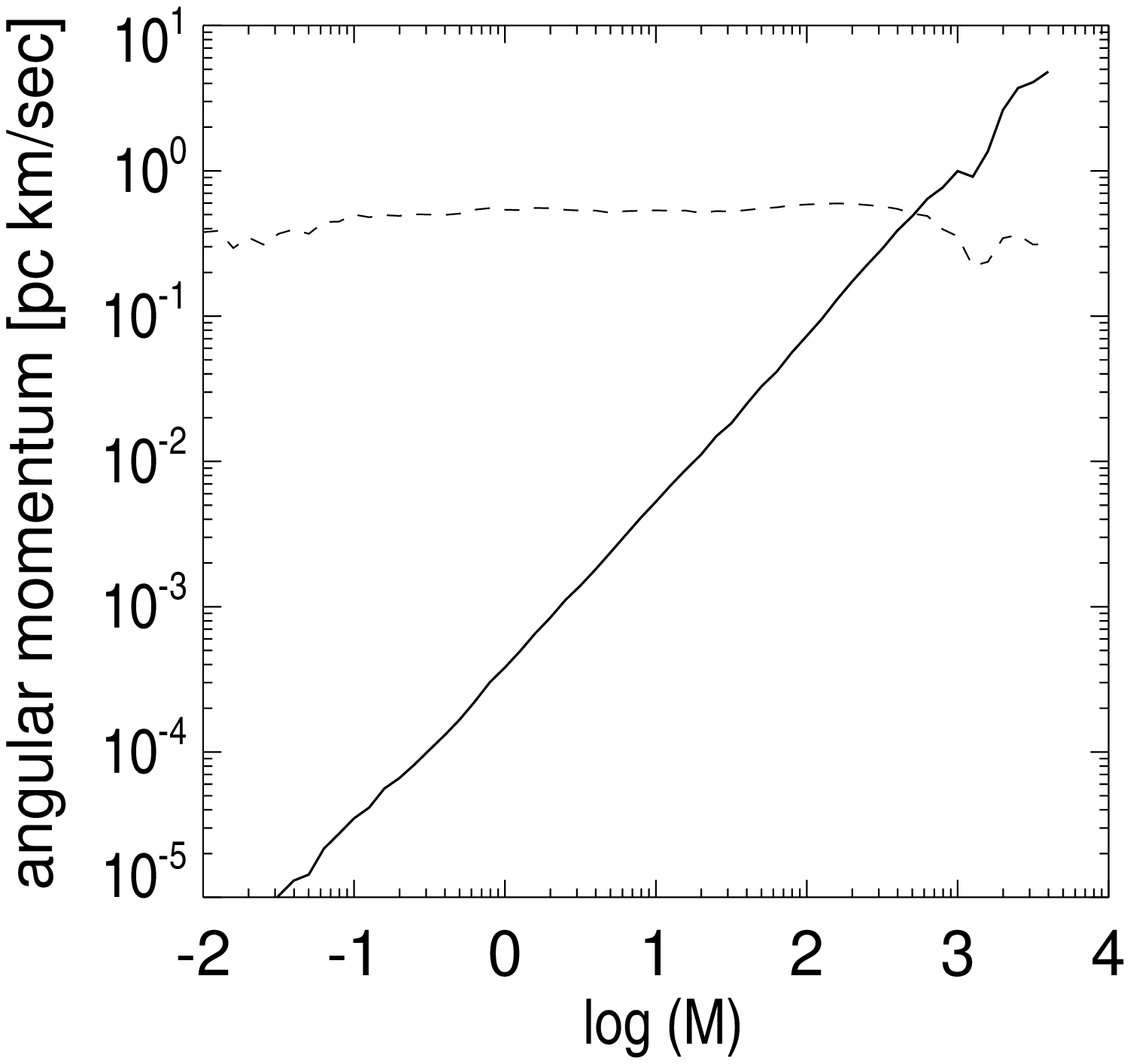}}
\caption{Specific angular momentum profile
in the gas cloud. The dashed line shows
the degree of rotational support as defined by equation
(\ref{eq:rot}). It is about 40-50\% for 
the entire cloud. 
\label{fig:AM}}
\end{inlinefigure}

In order to verify that numerical viscosity did not significantly
influence our results, we have explicitly checked the torque exerted
on the gas cloud.
In SPH, the equation of motion
for each gas particle is expressed as
\begin{equation}
m \frac{{\rm d} v}{{\rm d}t} = F_{\rm grav} + F_{\rm pressure} + F_{\rm visc} \, ,
\end{equation}
where the last term denotes the force owing to artificial viscosity 
(see Springel 2005 for exact expressions of each term).
Then the torque $N={\rm d}L/{\rm d}t$ exerted on a gas cloud consisting of $M$ particles
can be reduced to a sum of three components:
\begin{eqnarray}
N  &=& \sum_{i}^{M}  r_i \times m\dot{v}_i \nonumber \\
   &=& \sum_{i}^{M}  r_i \times (F_{\rm grav} + F_{\rm pressure} + F_{\rm visc})_i 
= N_{\rm grav} + N_{\rm pressure} + N_{\rm visc} \, .
\label{eq:torque}
\end{eqnarray}
We store the forces on each particle in output files and 
calculate explicitly all the terms in equation (\ref{eq:torque}).
We find that the dominant torques are $N_{\rm grav}$ and $N_{\rm pressure}$
throughout the evolution. Within the collapsing gas cloud 
($M_{\rm enc}< 200 M_{\odot}, r < 0.1 {\rm pc}$),
$N_{\rm grav}$ and $N_{\rm pressure}$ are completely dominant, and $N_{\rm visc}$ 
is always less than 10\% of $N_{\rm pressure}$ for all the mass shells.
Note that a small amount of dissipation by weak shocks is expected,
because there are some fluid elements that are moving at transonic velocities,
as found in the adaptive mesh refinement simulations of ABN02.
We have also performed the spherical collapse test of 
Norman, Wilson \& Barton (1980). We set up a rotating spherical
cloud as in Norman et al. (see also Truelove et al. 1998);
$M = 1 M_{\odot}, R=7.01 \times 10^{16} {\rm cm}$ 
and the initial rotation 
$\Omega = 3.04 \times 10^{-13} {\rm rad}\;{\rm sec}^{-1}$.
We use a half million particles to represent the sphere.
For an ideal (inviscid) fluid with no means of angular momentum redistribution,
the specific AM distribution is conserved, and it takes a simple
analytic form
\begin{equation}
M(<K) = M \left[ 1-\left(1-\frac{K}{\Omega R^2}\right)^{3/2}\right],
\end{equation} 
where $M(<K)$ is the mass of fluid elements that have specific AM
less than K. We measure the distribution at the initial 
time and that after about one free-fall time. The AM distribution
is found to be conserved rather well, with the deviation at the smallest
mass scale being less than 20\%. This is a similar level
of agreement as found in,
for example, the calculation of Truelove et al. (1998). 
Although a good level of AM conservation in this test
is expected for a Lagrangian hydrodynamics code, our test results
reassure that the artificial viscosity does not do any harm to
angular momentum transport in dynamically collapsing
gas spheres such as those studied in the present paper.


\begin{inlinefigure}
\resizebox{8cm}{!}{\includegraphics{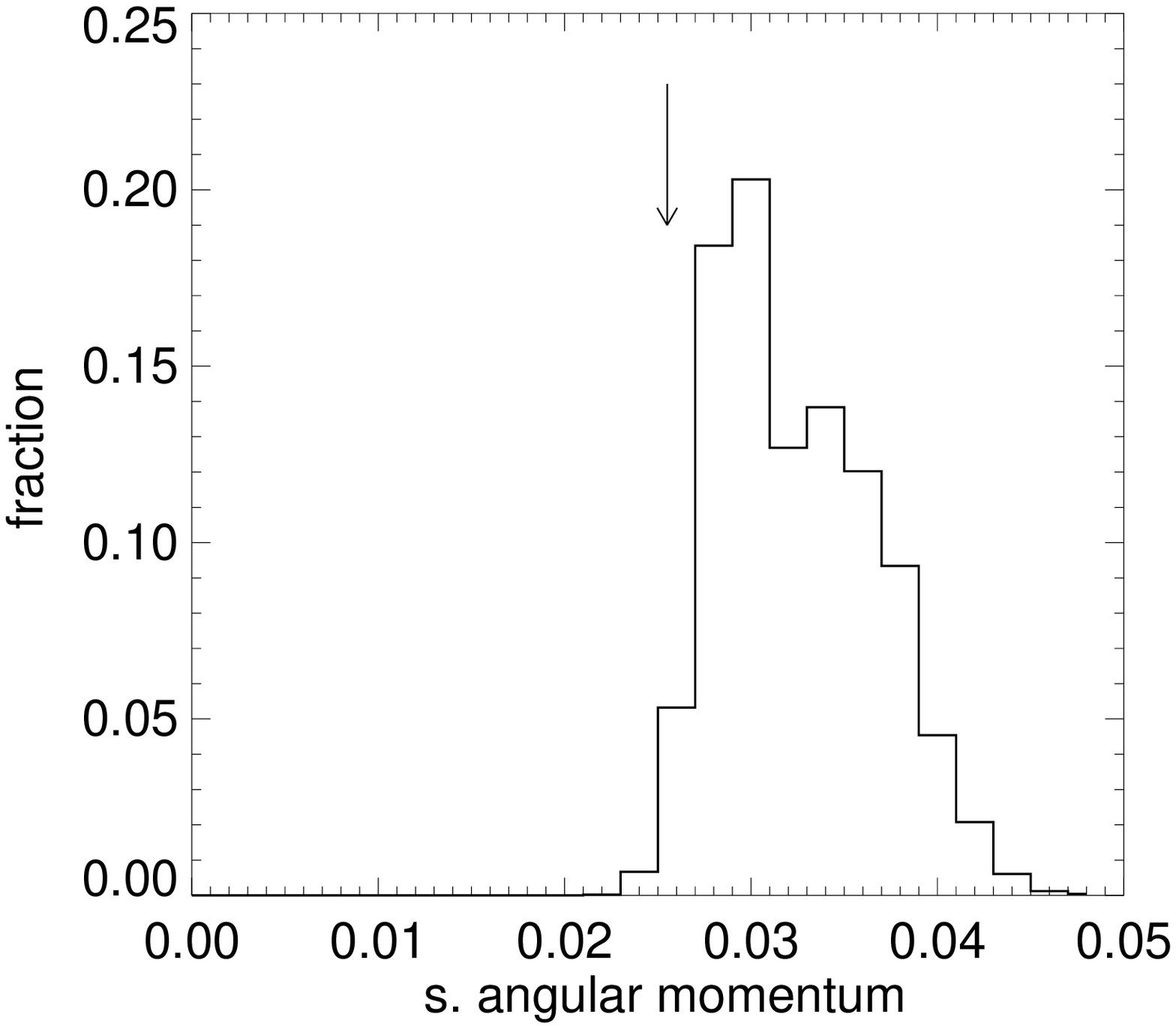}}
\resizebox{8cm}{!}{\includegraphics{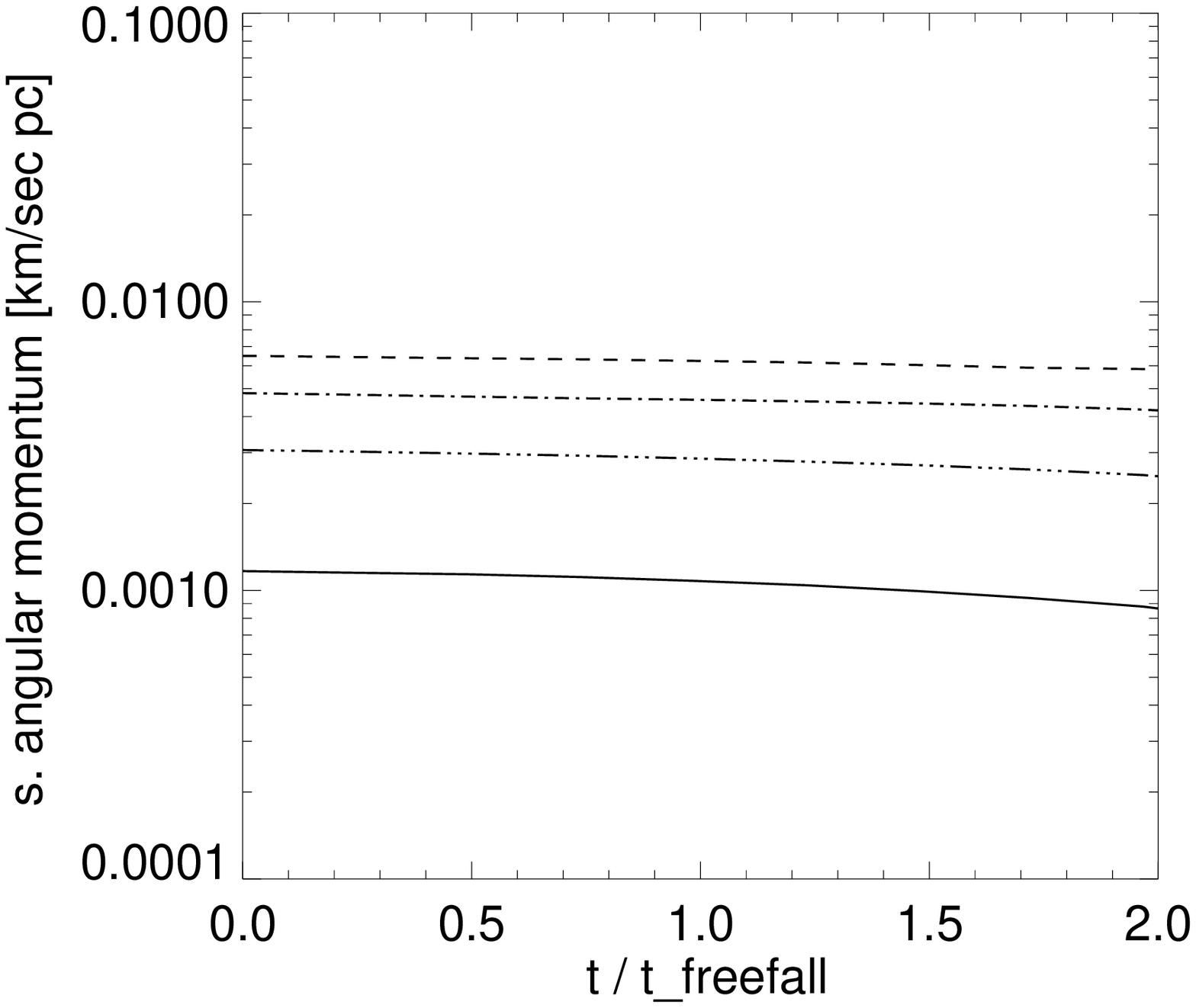}}
\caption{(Left) Probability distribution of specific angular momentum 
in the core region when the first collapse is being triggered.
The part that becomes the central region in the end
has a small angular momentum initially (indicated by the solid arrow),
whereas the temporal center (most dense region) before the collapse
has a larger angular momentum of $\sim 0.03$ pc km/sec.
(Right) Evolution of the `Lagrangian' specific
angular momentum with respect to the marked center
of the gas cloud for four groups of gas elements 
that are labeled according to their final mass coordinates.
\label{fig:AMevo}}
\end{inlinefigure}

\subsection{Evolution of the protostar}
It is still beyond the capability of the current simulation 
code and architecture to directly follow the evolution of the protostar
to the point where it settles onto the zero-age main sequence (ZAMS).
We tackle this demanding problem by employing the scheme for 
protostellar evolution of Stahler, Shu, \& Taam (1980a,b) as modified by 
Stahler et al. (1986a,b) and Omukai \& Palla (2001; 2003).
We assume that the gas accretion takes place in an approximately 
spherically symmetric manner. 
We use the mass accretion rate shown in
Fig. \ref{fig:massacc} as an input
to the protostellar evolution code.
While this assumption may seem over-simplified,
the gas mass accretion rate from our cosmological
simulation provides a reasonable estimate for true 
accretion rates onto a protostar, at least for the
initial accretion phase,
because the central gas cloud is roughly spherical and 
the gas in the envelope is not yet in a disk structure.
Moreover, the protostellar evolution calculations include various physics 
and thus will give a more self-consistent solution of the evolution
of a protostar for a given mass accretion rate.
We thus employ this method rather than 
use simple time-scale arguments as done
by ABN02 and Bromm \& Loeb (2004).

Briefly, the evolution of a protostar is treated as 
a sequence of a growing hydrostatic core with an accreting envelope.
The core is assumed to be in hydrostatic equilibrium, 
and the ordinary stellar structure equations are applied.
The structure of the accreting envelope is calculated with the assumption 
that the flow is steady for a given mass accretion rate. 
Only the inner envelope that is optically thick to continuum
is included. The effects of radiation pressure are considered in constructing 
the structure of the envelope.
The details of the calculation are found in Stahler, 
Palla, \& Salpeter (1986a,b) and Omukai \& Palla (2001; 2003).

We use the time-(equivalently, protostellar mass-)dependent 
accretion rate $\dot{M}_{\rm fid}$ derived from our simulation 
(Fig. \ref{fig:massacc}).
We refer to this rate as the fiducial rate hereafter. 
The evolution of the protostellar radius is shown in Fig. \ref{fig:proto_evo}.
as a function of the protostellar mass.
Considering the possibility that some fraction of infalling matter 
stacks in the circumstellar disk or is ejected as an outflow,
we also show cases of accretion rates reduced from our fiducial value
by a factor of 2/3 and 1/3. 

Initially, the Kelvin-Helmholz (KH) time $t_{\rm KH}$, which is 
the timescale for a star to lose its thermal energy by radiation, 
is longer than the accretion timescale $t_{\rm acc}$ because the 
temperature of the protostar is low and 
its opacity owing to free-free absorption is large 
($\kappa_{\rm ff}\propto T^{-7/2}$).
The accreted material simply piles up on the stellar surface 
without cooling.
This phase is called the adiabatic accretion phase and lasts 
up to $M_{\ast} \sim 10M_{\sun}$. 
During this phase, the radius first increases linearly (in log scale) 
with the protostellar mass, and then remains almost constant 
at $100R_{\sun}$.
This behavior can be explained as follows.

The protostellar radius in the adiabatic accretion phase 
depends on the protostellar mass and accretion rate as 
(Stahler et al. 1986a,b)
\begin{equation}
R_{\ast} \propto \dot{M_{\ast}}^{0.4} M_{\ast}^{0.3}.
\end{equation}
The increase of the radius in the earliest phase of evolution is explained by 
the increasing mass and accretion rate shown in Fig. \ref{fig:massacc}.
This is partly because our mass resolution is limited, and hence the output time
is slightly before the formation of an accretion shock.
Also, adjustment from a somewhat arbitrary initial structure 
to the appropriate accretion phase produces somewhat complex behavior
at $M_* < 0.3 M_{\odot}$. The accretion rate decays with mass as  
$\dot{M_{\ast}} \propto M_{\ast}^{-0.8}$ 
in the range $1M_{\odot} < M_{\ast} < 10M_{\odot}$ in our fiducial model.
Substituting this into the above relation, we see that the mass dependence 
of the radius almost cancels.
This explains the approximate constancy of the radius in this range.

\begin{inlinefigure}
\resizebox{10cm}{!}{\includegraphics{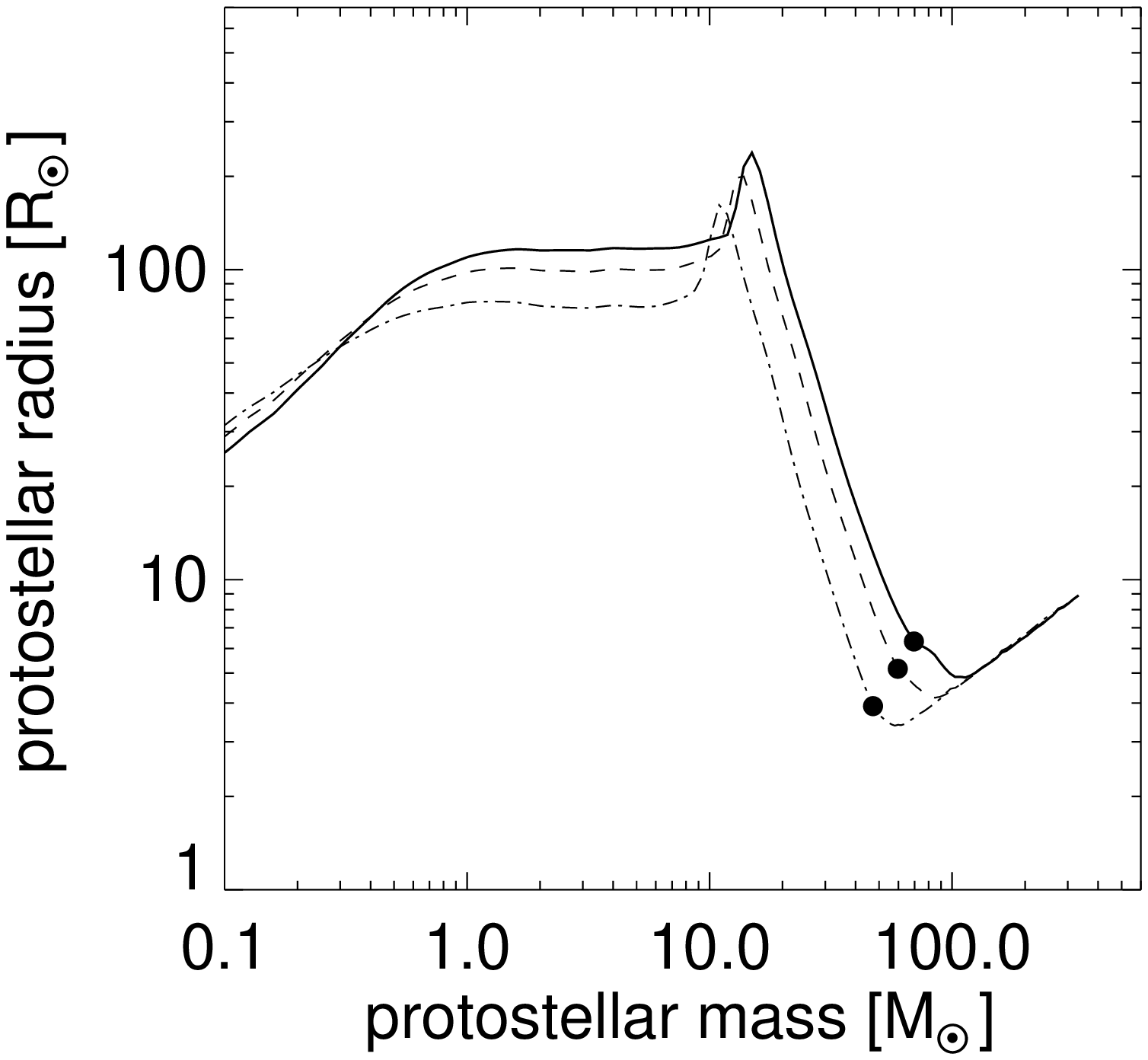}}
\caption{Proto-stellar evolution. The assumed accretion
rates are taken from our fiducial model $\dot{M}_{\rm fid}$ (solid),
 $(2/3) \dot{M}_{\rm fid}$ (dashed), $(1/3) \dot{M}_{\rm fid}$ (dash-dotted).
The solid points indicate the time when hydrogen burning begins.
\label{fig:proto_evo}}
\end{inlinefigure}

With increasing protostellar mass, increasing temperature in the stellar 
interior makes the opacity lower and the radiation diffuses out more
easily.
When the protostellar mass reaches $\sim 10M_{\odot}$, heat 
deposited in the 
interior is transported outwards by a luminosity wave, whose arrival to the 
surface makes the star swell suddenly (Stahler et al. 1986a,b). 
After that, the protostar enters the Kelvin-Helmholtz contraction 
phase. By losing thermal energy to radiation, the protostar shrinks 
as (Omukai \& Palla 2003)
\begin{equation}
R_{\ast} \propto \dot{M_{\ast}} M_{\ast}^{-2}.
\end{equation} 
The interior temperature increases until the contraction is 
halted by nuclear burning.

Energy generation by the p-p chain is not sufficient to 
stop the contraction of the already massive star.
When the central temperature reaches $10^{8}$K, 
hydrogen burning by the CNO cycle begins
with a slight amount of C synthesized by He burning. 
The onset of hydrogen burning by the CNO cycle 
is marked by a solid circle in the figure. 
The energy generation by hydrogen burning halts
contraction around $100 M_{\odot}$, 
and the star reaches the ZAMS.
The protostar relaxes to a ZAMS star eventually
within about $10^5$ years from the birth of the protostellar 
seed. In our calculation, there is no sign that accretion is
halted
by stellar feedback although the radiation pressure onto the 
accreting matter is included.
Even after its arrival onto the ZAMS, the star continues 
to accrete.
In principle, the star can grow in mass during its entire lifetime.
(Note that the accretion rate at the outer boundary is given
{\it as a boundary condition}. The gas accretion is sustained
because of this condition. See the discussion below.)
The stellar mass would become greater than a few hundred
solar masses during its main sequence lifetime
($\sim 3 \times 10^{6}$ yrs) if we were to allow mass accretion
to continue using our estimated accretion rate.
This mass scale is similar to the total gas mass of the dense core.
This coincidence is not by chance: since the stellar lifetime is longer 
than the free-fall time of the dense core ($\sim 3 \times 10^{5}$ yrs for 
$n_{\rm H} \sim 10^{4}{\rm cm^{-3}}$), the whole core falls onto the star.
  
In order for the accretion to continue, the total luminosity of the protostar 
$L_{\rm tot}$, given by the sum of the luminosity from the stellar interior $L_{\ast}$
and that from the accretion shock $L_{\rm acc}=GM_{\ast} \dot{M}_{\ast}/ R_{\ast}$, 
must not exceed the Eddington limit 
$L_{\rm Edd}=4 \pi c G M_{\ast}/ \kappa_{\rm es}$, 
where $\kappa_{\rm es}$ is the electron scattering opacity, 
the dominant source of opacity in the inner envelope.
The total luminosity approaches the Eddington limit as the star approaches 
the ZAMS, decreasing the radius and increasing the luminosity.
To reach the ZAMS, the mass accretion rate must be less than
\begin{equation}
\dot{M}_{\rm crit} \simeq \frac{(L_{\rm Edd}-L_{\rm ZAMS}) R_{\rm ZAMS}}{G M_*},
\label{eq:critacc}
\end{equation}
where $R_{\rm ZAMS}$ and $L_{\rm ZAMS}$ are the radius and luminosity of 
the ZAMS star of mass $M_{\ast}$.
The dependence of the critical accretion rate on stellar mass 
is very weak and becomes approximately constant at
$\dot{M}_{\rm crit}\simeq 4\times 10^{-3} M_{\odot} {\rm yr}^{-1}$
(Omukai \& Palla 2003).
This means that once the star reaches the ZAMS, the luminosity remains always 
below the Eddington limit and the accretion continues unimpeded.
In Fig. \ref{fig:massacc}, we show the critical rate by the dotted line. 
Although the accretion rate is higher than the critical value
in the early evolution, the total luminosity 
falls below the Eddington limit owing to the larger radius and 
smaller interior luminosity in that phase than a ZAMS star of 
the same mass.
For the fiducial accretion rate, the accretion rate drops below 
the critical value at $M_{\ast} \simeq 30 M_{\odot}$, well before 
the star reaches the ZAMS.
Consequently, the total luminosity never exceeds the Eddington limit 
and accretion continues.

We also explored two other cases besides our fiducial
model. For these cases, we attempt to model crudely 
the feedback effects from the protostar 
and the consequences of rotation by reducing the mass 
accretion rate from the original value.
The formed protostar could launch a protostellar outflow 
if a small magnetic field is present initially (Machida et al. 2006).
Some fraction of infalling matter in the envelope is eventually 
ejected as an outflow without falling onto the central star and the 
accretion rate onto the protostar is reduced.
To this end, we use reduced accretion rates of $(2/3) \dot{M}_{\rm fid}$
and $(1/3) \dot{M}_{\rm fid}$. The overall evolution of these cases
is similar to that of the fiducial case, but characteristic mass scales are
shifted to somewhat lower values. 
For the case with $\dot{M} = (1/3) \dot{M}_{\rm fid}$,
the protostar reaches the ZAMS when its mass is $\sim 60 M_{\odot}$.
In the case of disk accretion, not only the accretion rate, but also 
the surface boundary condition of the star is altered.
For smaller accretion rates ($10^{-5}-10^{-4}M_{\odot}/{\rm yr}$), 
Palla \& Stahler (1992) studied the cases of different boundary 
conditions, which correspond to spherical or disk accretion cases.
Intriguingly, they showed that the evolutionary behavior is quite similar 
in both cases qualitatively; the star reaches ZAMS somewhat earlier
in the disk accretion case.

In summary, the accretion is not halted by radiative feedback
for the fiducial and reduced accretion rates.
Note, however, that we only considered the inner envelope, 
which is optically thick to continuum.
In the outer envelope, the Ly$\alpha$ opacity could be important
for halting accretion (Doroshkevich \& Kolesnik 1976; Tan \& McKee 2004). 
The expansion of an \HII region might be an another possibility to hinder 
accretion (Larson \& Starrfield 1971; Tan \& McKee 2004).
Since the expansion of an \HII region is quenched 
for a large accretion rate assuming spherical symmetry (e.g., Omukai \& Inutsuka 2002),
non-spherical effects must be included for evaluating the exact time 
behavior of the expansion of the \HII region.
While these uncertainties remain in determining the exact mass-scale of 
the first stars, the fact that they are rather massive 
seems to be robust.

\section{Summary and Discussion}
We have developed numerical techniques to implement the required
chemical and radiative processes for following the thermal 
evolution of primordial gases to very high densities.
Using a cosmological simulation with hydrodynamics and chemistry, 
we have studied the formation and evolution of pre-stellar 
gas clouds in a $\Lambda$CDM universe. 
Our three-dimensional simulation for the first time
resolves the inner $\sim 1M_{\odot}$ fully molecular core,
allowing us to obtain correct density, temperature, and velocity
structure around the primordial protostar. These quantities altogether
yield an accurate gas mass accretion rate
around the protostar that was uncertain in previous works.
We have derived three important facts from our simulations: 
(1) the primordial gas cloud in a cosmological minihalo 
does not fragment, but yields a single proto-stellar seed,
(2) the cloud core has a small angular momentum
so that rotation does not halt the collapse, nor is a disk
formed, and 
(3) the rate of accretion from the cloud envelope is large. 
From these facts, we conclude that the first stars are massive. 
By performing a proto-stellar evolution calculation, 
we derive that the protostar grows to $\sim 100 M_{\odot}$ for the particular
case found in our simulation. We derive the stellar mass
for the first time from a self-consistent calculation
of proto-stellar evolution, coupled with a cosmological simulation.

While we simulated only a single halo in detail,
the particular case shows a convincing and plausible case
for the formation of massive stars 
in early mini-halos in the CDM model.
The fate of such massive primordial stars (Population III),
whether or not they are commonly born in various environments, 
are of considerable interest. 
They are likely responsible for various feedback effects in the early Universe
(Haiman, Rees, \& Loeb 1997; Bromm, Yoshida, \& Hernquist 2003; Yoshida,
Bromm, \& Hernquist 2004; see Ciardi \& Ferrara 2005 for a review). 
Stars with mass $140-260 M_{\odot}$
are believed to trigger pair-instability supernovae
and expel all the processed heavy elements
in the explosion (Bond, Arnett, \& Carr 1984; Heger \& Woosley 2002). 
It has also been suggested that less massive stars  
die as hypernovae (Umeda \& Nomoto 2002),
producing a quite different metal yield from the
pair-instability case. Interestingly, the observed abundance
pattern of extremely metal poor stars appears
to favor the latter scenario (Umeda \& Nomoto 2003; 
Frebel et al. 2005). 
Energetic supernovae are also destructive,
being effective in evacuating the halo
gas (Bromm, Yoshida \& Hernquist 2003; Wada \& Venkatesan 2003;
Kitayama \& Yoshida 2005) and possibly quenching 
further star-formation in that region.
A single supernova would suffice 
to pollute a large volume of the interstellar
matter from which very metal-poor stars might
have been formed. Metal-enrichment by the first supernovae 
may also lead to the formation of ordinary stellar populations including
low-mass stars (e.g. Mackey et al. 2003) if metal-enrichment
is relatively confined within a small region.
Intriguingly, Jimenez \& Haiman (2006)
argue that metal-mixing could be inefficient and
that primordial stars are formed in galaxies at $z<5$.

Massive stars emit a large number
of ionizing photons and can ionize
a large volume of the intergalactic medium
around them (Kitayama et al. 2004; Whalen et al. 2004).
Recently, Yoshida (2006) carried out
radiation-hydrodynamic calculations of early \HII
regions and concluded that radiation
from a massive Population III star
can completely ionize the gas within the host
halo, quenching further gas condensation
and star-formation for a significant period of 
cosmic time. Both massive and very massive ($>300 M_{\odot}$) 
stars likely leave a black hole remnant.
It is plausible that the remnants become
the seeds for supermassive black holes
that power luminous quasars found at $z \ga 6$
(Li et al. 2006, in preparation) and subsequently
at lower redshifts (e.g. Di Matteo et al. 2005;
Hopkins et al. 2005, 2006).
Because bottom-up hierarchical 
structure formation is a generic prediction of the CDM model,
it is natural to think that the early population stars
affect and even set the initial conditions
for primeval galaxy formation.  It is important to 
explore feedback from the first stars in more detail
under a proper cosmological set-up.

The prospects for detecting the first generation of stars
appear promising. Supernovae resulting from the collapse
of massive stars can be observed if they trigger
gamma-ray bursts (GRBs). There is growing
evidence that long duration GRBs are associated with
energetic supernovae (e.g. Hjorth et al. 2003).
Gou et al. (2004) suggest that the afterglow of 
high-redshift GRBs can be detected in X-rays.
There are also a broad range of observations
through which we can, in principle, indirectly observe 
star-formation at high redshifts, through, for example,
angular fluctuations in redshifted 21 cm emission from
intergalactic gas as stars reionize the Universe (see,
e.g. Zaldarriaga et al. 2004; Furlanetto et al. 2004;
Zahn et al. 2006),
or from secondary anisotropies imprinted from stellar
radiation on the CMB (e.g. McQuinn et al. 2005; Zahn
et al. 2005).

Finally, we remark that it is still too early to definitely conclude 
the exact mass or mass range of the first stars.
The actual mass accretion could take place
in a more complicated manner than we assume. Feedback from the protostar 
itself can affect the dynamics and thermal evolution of the accreting gas.
Ultra-violet radiation from the protostar likely affects the gas infall rate. 
However, primordial gas has a much smaller opacity because of the absence
of dust, and thus radiation pressure is generally not important (Omukai \& Inutsuka 2002).
The effect of radiation pressure can also be reduced if protostellar outflows 
provide optically thin cavities through which photons can escape (Krumholz, McKee \& Klein 2005). 
Future calculations of proto-stellar evolution
need to include these feedback effects 
onto the infalling gas. Also, accurate modeling of disk accretion
may be necessary in general cases (Tan \& McKee 2004).
Although our simulation is currently resolution-limited
(rather than physics-limited),
with the maximum density being still five orders of
magnitude smaller than stellar densities,
we foresee that it will be possible to carry out
an {\it ab initio} calculation of the formation of primordial 
protostars in the near future. 

\bigskip

NY thanks Fumitaka Nakamura, George Field, Simon Glover, Gao Liang,
and Brian O'Shea for insightful discussions.
The simulations were performed 
at the Center for Parallel Astrophysical Computing 
at Harvard-Smithsonian Center for Astrophysics, 
at the Center for Computational Cosmology at Nagoya University,
and at the Data-Reservoir
at the University of Tokyo. We thank Mary Inaba and Kei Hiraki
for providing the computing resource at U-Tokyo.
The work is supported in part by the Grants-in-Aid 
for Young Scientists (A) 17684008 (N.Y.), and 
by Young Scientists (B) 18740117, Scientific Research on Priority Areas 18026008
(K.O.) by the Ministry of Education, 
Culture, Science and Technology of Japan.
This work was supported in part by NSF grants
AST 03-07433 and AST 03-07690,
and NASA ATP grants NAG5-12140, NAG5-13292, and NAG5-13381.


\clearpage
\appendix
\begin{deluxetable}{llll}
\tablecolumns{4}
\tablecaption{REACTION RATE COEFFICIENTS}
\tablehead{
&\colhead{Reactions} & \colhead{Rate Coefficients
 (cm$^3$s$^{-1}$)} 
& \colhead{Reference}}
\startdata

  (1) & ${\rm H}       + e        \rightarrow {\rm H^+}     + 2e   $  
      &  $k_{1} = \exp [-32.71396786 
                   + 13.536556     \;(\ln T_{\rm e})   $ \\  & & ~~~~~~~$ 
		   - 5.73932875     \;(\ln T_{\rm e})^2 
		   + 1.56315498     \;(\ln T_{\rm e})^3 $ \\  & & ~~~~~~~$ 
		   - 0.2877056    \;(\ln T_{\rm e})^4
		   + 0.0348255977 \;(\ln T_{\rm e})^5 $ \\  & & ~~~~~~~$ 
		   - 0.00263197617   \;(\ln T_{\rm e})^6 
		   + 0.000111954395 \;(\ln T_{\rm e})^7 $ \\  & & ~~~~~~~$ 
		   - 2.03914985\times 10^{-6} \;(\ln T_{\rm e})^8]$  & 1 \\ 
\\
  (2) & ${\rm H^+}     + e        \rightarrow {\rm H}       + h\nu $ 
      & $k _{2} = \exp[-28.6130338
		   - 0.72411256    \;\ln T_{\rm e}       $ \\  & & ~~~~~~~$ 
		   - 0.0202604473   \;(\ln T_{\rm e})^2   
		   - 0.00238086188  \;(\ln T_{\rm e})^3   $ \\  & & ~~~~~~~$   
		   - 0.000321260521 \;(\ln T_{\rm e})^4   $ \\  & & ~~~~~~~$ 
		   - 0.0000142150291 \;(\ln T_{\rm e})^5   $ \\  & & ~~~~~~~$ 
		   + 4.98910892 \times 10^{-6} \;(\ln T_{\rm e})^6 $ \\  & & ~~~~~~~$ 
		   + 5.75561414 \times 10^{-7} \;(\ln T_{\rm e})^7 $ \\  & & ~~~~~~~$ 
		   - 1.85676704 \times 10^{-8} \;(\ln T_{\rm e})^8 $ \\  & & ~~~~~~~$ 
		   - 3.07113524 \times 10^{-9} \;(\ln T_{\rm e})^9] $ & 1 \\ 
\\ 
  (3) & ${\rm He}      + e        \rightarrow {\rm He^+}    + 2e   $   
      & $k_{3}= \exp[-44.09864886
		   + 23.91596563      \;\ln T_{\rm e}     $ \\  & & ~~~~~~~$ 
		   - 10.7532302      \;(\ln T_{\rm e})^2 
		   + 3.05803875      \;(\ln T_{\rm e})^3 $ \\  & & ~~~~~~~$ 
		   - 0.56851189  \;(\ln T_{\rm e})^4
		   + 0.0679539123 \;(\ln T_{\rm e})^5 $ \\  & & ~~~~~~~$ 
		   - 0.00500905610 \;(\ln T_{\rm e})^6
		   + 0.000206723616  \;(\ln T_{\rm e})^7 $ \\  & & ~~~~~~~$ 
		   - 3.64916141   \times 10^{-6}   \;(\ln T_{\rm e})^8]$  & 1 \\ 
\\
  (4) & ${\rm He^+}    + e        \rightarrow {\rm He}      + h\nu $ 
      & $k_{4r}=3.925\times 10^{-13} T_{\rm e}^{-0.6353}$ \\ & &  
      $ k_{4d}=1.544\times 10^{-9} T_{\rm e}^{-1.5} \exp(-48.596/T_{\rm e})
	\times [0.3 + \exp(8.10/T_{\rm e})]$ & 1 \\
\\
  (5) & ${\rm He^+}    + e        \rightarrow {\rm He^{++}} + 2e   $   
      & $k _{5}=\exp[-68.71040990
	 	   +  43.93347633      \;\ln T_{\rm e}       $ \\  & & ~~~~~~~$ 
		   -  18.4806699      \;(\ln T_{\rm e})^2
		   +   4.70162649   \;(\ln T_{\rm e})^3   $ \\  & & ~~~~~~~$ 
		   -   0.76924663     \;(\ln T_{\rm e})^4
		   +   0.08113042    \;(\ln T_{\rm e})^5   $ \\  & & ~~~~~~~$ 
		   -   0.00532402063  \;(\ln T_{\rm e})^6   $ \\  & & ~~~~~~~$ 
		   +   0.000197570531  \;(\ln T_{\rm e})^7   $ \\  & & ~~~~~~~$ 
		   -   3.16558106 \times 10^{-6}   \;(\ln T_{\rm e})^8)$  & 1 \\
\\
  (6) & ${\rm He^{++}} + e        \rightarrow {\rm He^+}    + h\nu $ 
      & $k_{6} = 2\times k_{2}(T_{\rm e}/4)$ & 1 \\
\\
  (7) & ${\rm H}       + e        \rightarrow {\rm H^-}     + h\nu $ 
      & $k_{7} = 1.4\times 10^{-18}T^{0.928} \exp(-T/16200)$ & 2 \\
\\
  (8) & ${\rm H^{-}}   + {\rm H}     \rightarrow \rm H_{2} + e   $ 
      & $k_{8} = 4.0\times 10^{-9}T^{-0.17}$ & 2 \\
 \\
  (9) & ${\rm H}       + {\rm H^{+}} \rightarrow {\rm H_{2}^{+}} + h\nu   $ 
      & $k_{9}={\rm dex} [-19.38 -1.523\log T+1.118(\log T)^2 -0.1269(\log T)^3]$  & 2,3,4\\
\\
  (10) & ${\rm H_{2}^{+}} + {\rm H}  \rightarrow {\rm H_{2}^{*}} + {\rm H^{+}}   $ 
       & $k _{10} = 6.0\times 10^{-10}$ & 5 \\
\\
  (11) & ${\rm H_{2}}  + {\rm H}     \rightarrow {\rm 3H}                         $ 
       & fit by reference 6 & 6 \\
\\
  (12) & ${\rm H_{2}}  + {\rm H^{+}} \rightarrow {\rm H_{2}^{+}} + {\rm H}       $ 
       & $k_{12} = \exp(-21237.15/T) \times [
	- 3.3232183\times 10^{-7}   $ \\  & & ~~~~~~~$           
	+ 3.3735382\times 10^{-7}\;(\ln T     
	- 1.4491368\times 10^{-7}\;(\ln T)^2   $ \\  & & ~~~~~~~$ 
	+ 3.4172805\times 10^{-8}\;(\ln T)^3     
	- 4.7813720\times 10^{-9}\;(\ln T)^4   $ \\  & & ~~~~~~~$ 
	+ 3.9731542\times 10^{-10}\;(\ln T)^5    
        - 1.8171411\times 10^{-11}\;(\ln T)^6   $ \\  & & ~~~~~~~$ 
	+ 3.5311932\times 10^{-13}\;(\ln T)^7]
      $ & 7 \\
\enddata
\label{tab:chem rate1}
\end{deluxetable}

\begin{deluxetable}{llll}

\tablecolumns{4}
\tablecaption{REACTION RATE COEFFICIENTS (continued)}
\tablehead{
&\colhead{Reactions} & \colhead{Rate Coefficients
 (cm$^3$s$^{-1}$)} 
& \colhead{Reference}}
\startdata
  (13) & ${\rm H_{2}}  + e           \rightarrow {\rm 2H}        + e             $ 
       & $k_{13} = 3.73\times 10^{-9}\; T^{0.1121} \exp(-99430/T)$  & 9 \\
\\
  (14) & ${\rm He^{+}}  + {\rm H}    \rightarrow {\rm He}        + {\rm H^{+}} + h\nu  $ 
       & $k_{14} = 1.20 \times 10^{-15} (T/300)^{0.25}$ & 5, 10 \\
\\
  (15) & ${\rm He}      + {\rm H^{+}} \rightarrow {\rm He^{+}}   + {\rm H}       $ 
       & $k_{15} = 1.26 \times 10^{-9} T^{-0.75} \exp(-127500/T)$ ~~~($T<10^4$ K) \\  
       & 
       & $k_{15} =  4 \times 10^{-37} T^{4.74}$ ~~~~~~~~~~~~~~~~~~~~~~~~~~~~~~($T>10^4$ K) & 5, 11 \\
\\
  (16) & ${\rm H^{-}}   + e           \rightarrow {\rm H}        + 2e            $ 
       & $k_{16} = \exp [-18.01849334
		    + 2.3608522     \;\ln T_{\rm e}   $ \\  & & ~~~~~~~$ 
		    - 0.28274430    \;(\ln T_{\rm e})^2 
		    + 0.0162331664   \;(\ln T_{\rm e})^3   $ \\  & & ~~~~~~~$ 
		    - 0.0336501203   \;(\ln T_{\rm e})^4
		    + 0.0117832978   \;(\ln T_{\rm e})^5   $ \\  & & ~~~~~~~$ 
		    - 0.00165619470  \;(\ln T_{\rm e})^6
		    + 0.000106827520 \;(\ln T_{\rm e})^7    $ \\  & & ~~~~~~~$ 
		    - 2.63128581     \times 10^{-6}  \;(\ln T_{\rm e})^8]$ & 1 \\
\\
  (17) & ${\rm H^{-}}   + {\rm H}     \rightarrow {\rm 2H}       + e             $ 
       & $k_{17} = \exp [-20.37260896
		    + 1.13944933  \;\ln T_{\rm e}      $ \\  & & ~~~~~~~$ 
		    - 0.14210135 \;(\ln T_{\rm e})^2
		    + 0.0084644554   \;(\ln T_{\rm e})^3  $ \\  & & ~~~~~~~$ 
		    - 0.0014327641 \;(\ln T_{\rm e})^4
		    + 0.00020122503 \;(\ln T_{\rm e})^5  $ \\  & & ~~~~~~~$ 
		    + 0.000086639632 \;(\ln T_{\rm e})^6  $ \\  & & ~~~~~~~$ 
		    - 0.000025850097 \;(\ln T_{\rm e})^7  $ \\  & & ~~~~~~~$ 
		    + 2.4555012 \times 10^{-6} \;(\ln T_{\rm e})^8 $ \\  & & ~~~~~~~$ 
		    - 8.0683825 \times 10^{-8}   \;(\ln T_{\rm e})^9]$ & 1 \\
\\
  (18) & ${\rm H^{-}}   + {\rm H^{+}} \rightarrow {\rm 2H}                       $
& $k_{18}=6.3\times 10^{-8} + 5.7 \times 10^{-6}T^{-0.5} -9.2\times 10^{-11}T^{0.5}
+ 4.4\times 10^{-13}T$ 
       & 2, 12 \\
\\
  (19) & ${\rm H^{-}}   + {\rm H^{+}} \rightarrow {\rm H_{2}^{+}} + e            $ 
       & $ k_{19} = 4.0\times 10^{-4} T^{-1.4} \exp(-15100.0/T)$ ~~~($T>10^4$ K) & \\ 
       &        
       & $ k_{19} = 1.0\times 10^{-8} T^{-0.4}$  ~~~~~~~~~~~~~~~~~~~~~~~~~~($T<10^4$ K) & 13 \\
\\		   
  (20) & ${\rm H_{2}^{+}} + e         \rightarrow {\rm 2H}                       $ 
       & $k_{20} = 2.0\times 10^{-7} T^{-0.5}$ & 2 \\
\\
  (21) & ${\rm H_{2}^{+}} + {\rm H^{-}} \rightarrow {\rm H}      +  {\rm H_{2}}  $ 
     & $k_{21} = 5.0\times 10^{-7} (T/100)^{-0.5}$ & 1 \\

\\
  (22) & ${\rm 3H} \rightarrow {\rm H_{2}}      +  {\rm H}  $ 
     & $k_{22} = 5\times 10^{-29} T^{-1}$ & 8 \\
\\
  (23) & ${\rm 2H} + {\rm H_{2}} \rightarrow 2{\rm H_{2}} $ 
     & $k_{23} = k_{22}/8$ & 8 \\
\\
  (24) & ${\rm H_{2}}  +  {\rm H_{2}} \rightarrow {\rm 2H} + {\rm H_{2}} $ 
       & $k_{24} = 8.125\times 10^{-8} T^{-1/2} \exp(-52000.0/T) $ \\  & & ~~~~~~~$ 
         \times(1.0-\exp (-6000.0/T))$ & 8 \\

\tablecomments{
We denote $T_{\rm e}$ for $T$ in eV, otherwise $T$ is in K. References are
(1): Abel et al. (1997); 
(2): Galli \& Palla (1998);
(3): Ramaker \& Peek (1976);
(4): Stancil et al. (1993);
(5): Glover \& Brand (2003);
(6): Martin et al. (1996);
(7): Savin et al. (2004a,b);
(8): Palla, Stahler, \& Salpeter (1983);
(9): Stibbe \& Tennyson (1999);
(10): Zygelman et al. (1989);
(11): Kimura et al. (1993);
(12): Peterson et al. (1971);
(13): Shapiro \& Kang (1987)
}

\enddata
\label{tab:chem rate2}
\end{deluxetable}

\end{document}